\documentclass[twocolumn,pre,superscriptaddress]{revtex4-1}
\usepackage{amsfonts}
\usepackage{amsmath}
\usepackage{amssymb}
\usepackage{color}
\usepackage{graphicx}
\usepackage{bm}
\usepackage{esint}
\usepackage{ulem}
\usepackage{xcolor}

\usepackage{soul}
\usepackage{cancel}
\usepackage{braket}

\begin{document}

\title{Transmission of waves through a pinned elastic {medium}}

\date{\today}

\author{Tsuyoshi Yamamoto}
\affiliation{Institute for Solid State Physics, the University of Tokyo, Kashiwa, Chiba 277-8581, Japan}
\author{Leonid I. Glazman}
\affiliation{Departments of Physics and Applied Physics, Yale University, New Haven, CT 06520, USA}
\author{Manuel Houzet}
\affiliation{Univ.~Grenoble Alpes, CEA, Grenoble INP, IRIG, PHELIQS, F-38000 Grenoble, France}

\begin{abstract}

We investigate the scattering of elastic  waves off a disordered region described by a one-dimensional random-phase sine-Gordon model. The collective pinning results in an effective static disorder potential with universal and non-Gaussian correlations, acting on propagating waves.
We find signatures of the correlations in the wave transmission in a wide frequency range, which covers both the weak and strong localization regimes.
{Our theory elucidates the dynamics of collectively-pinned phases occurring in any natural or synthetic elastic medium. The latter one is exemplified by a one-dimensional array of Josephson junctions, for which we specify our results. The obtained results provide benchmarks for the array-enabled quantum simulations addressing the dynamics in  broader and yet-unexplored domains of  individual pinning and quantum Bose glass.}

\end{abstract}

\maketitle

\section{Introduction}

The interplay of elasticity and disorder has attracted a wide interest because of its relevance to describe a large variety of physical systems, both classical (superconducting vortex lattice~\cite{Larkin1970}, ferroelectric and magnetic domain wall~\cite{Imry1975}, charge density wave~\cite{Gorkov1977,Fukuyama1978}) and quantum (Bose glass~\cite{Giamarchi1988,Fisher1989}, Wigner crystals~\cite{Fukuyama1978b,Chitra1998}). It has been realized that a weak disorder potential destroys long-range order in any dimension $D<4$ due to the collective pinning mechanism first described by Larkin in the 1970s~\cite{Larkin1970}. Since then, many technical tools have been developed to {treat} the correlation functions in disordered elastic media both in space~\cite{Giamarchi1995} and time~\cite{Giamarchi1996}, and use them to unveil a glassy behavior in the melting and creep of the charge density, or discuss the bulk electromagnetic absorption, see Ref.~\cite{Giamarchi2009} for a review. Nevertheless, the interplay of elasticity and disorder remains a difficult optimization problem, with many remaining open questions both for static and dynamic properties.

Applied to the charge density waves' propagation, the classical dynamics can be analyzed by considering the properties of small oscillations on top of a static charge density~\cite{Gorkov1977}. In the weak pinning theory, the static charge density determines a correlated, non-Gaussian disorder potential in the wave equation for plasmons. {A one-dimensional (1D) nature of the system allows for most comprehensive analytical and numerical studies.} So far, both the spatially-averaged~\cite{Fukuyama1978,Feigelman1980,Feigelman1981,Aleiner1994,Fogler2002,Gurarie2002,Gurarie2003} and local~\cite{Houzet2019} plasmon density of states {in a 1D array} have been computed. The novelty of our work is to consider the plasmon transmission across a disordered region of a finite length. This is the standard quantity for revealing the localization properties of the disorder~\cite{Lifshitz1988,Pendry1994,Makarov2012}. We find signatures of the collective pinning in a wide frequency range that extends from the Anderson localization limit at large frequencies to the limit of strong localization at low frequencies.

\begin{figure}
\includegraphics[width=0.9\columnwidth]{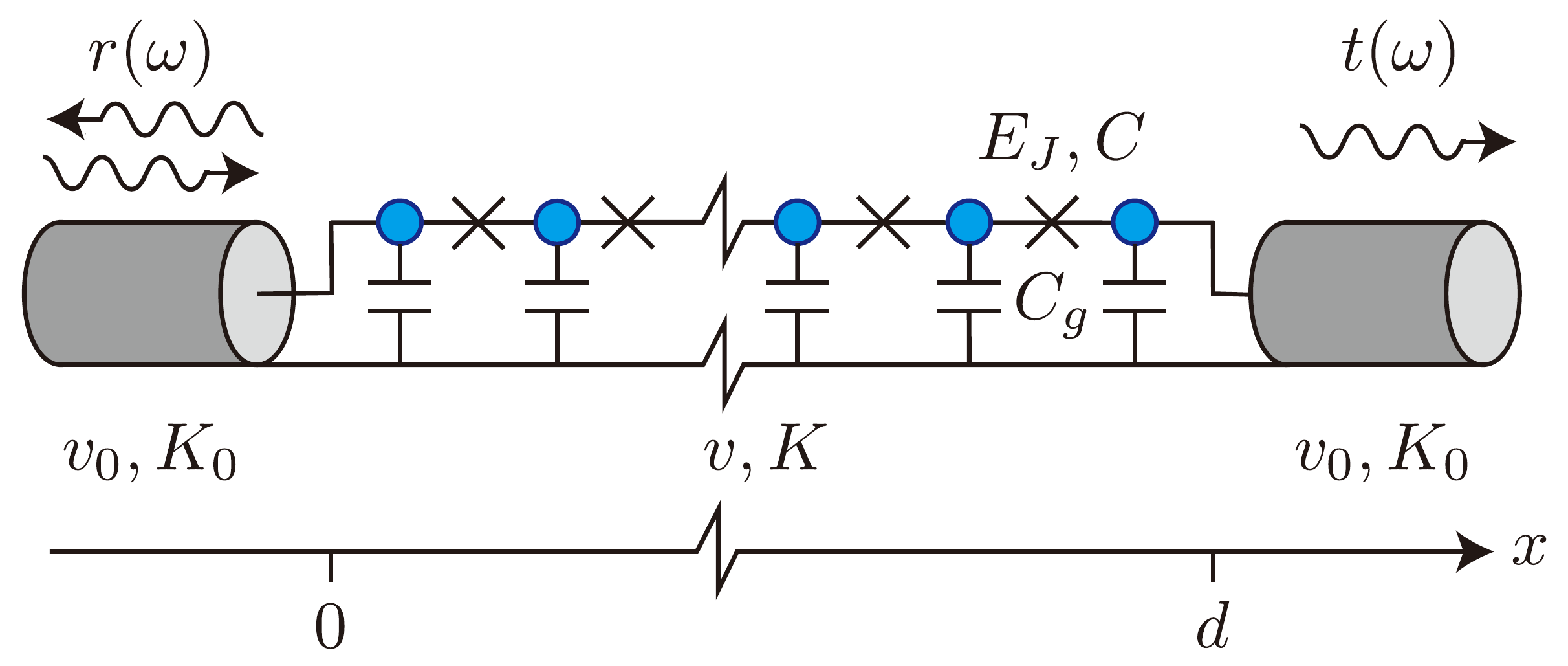}
\caption{\label{fig:setup}
As an example of wave propagation through a disordered elastic medium, we consider the microwave plasmon scattering by a one-dimensional Josephson-junction array of length $d$. The Josephson energy and capacitance of each junction along the array are $E_J$ and $C$, respectively; the ground capacitance of each superconducting island is $C_g$. In the presence of random background charges, the Josephson-junction array realizes a random interacting medium. Classically, an incident plasmon wave from one of the waveguides contacted to the array is either reflected to the same waveguide or transmitted to the opposite waveguide, with the respective amplitudes $r(\omega)$ and $t(\omega)$ at frequency $\omega$. The plasmon velocities in the waveguides and the array are, respectively, $v_0$ and $v$; the waveguide and array impedances are, respectively, $R_Q/2K_0$ and $R_Q/2K$, where $R_Q=\pi \hbar/2e^2$ is the resistance quantum.
}
\end{figure}

In the collective pinning regime, the Larkin length is the natural length scale for the correlation of the static disorder induced in the pinned medium{~\cite{Larkin1970,Imry1975,Fukuyama1978}}. We find that the universality in the static correlation gives rise to universality of the finite-frequency wave scattering by the pinned medium. While the scattering properties are universal, they differ from those for a Gaussian white{-}noise disorder. The differences are most prominent both in the limits of almost-ballistic propagation and of strongly localized regime. 

At sufficiently high frequencies, i.e., when the frequency-dependent mean free path exceeds the length of the system, the differences are encoded in the forward-scattering amplitude. This amplitude is sensitive to the static disorder correlations, in spite of the wavelength being shorter than the Larkin length. 

At low frequencies, corresponding to wavelengths larger than the Larkin length, we focus on the disorder-averaged transmission coefficient $\Braket{T}$ and its logarithm (commonly referred to as the Lyapunov exponent). Previously these two characteristics, $\Braket{T}$ and $\Braket{\ln T}$, were studied extensively for a Gaussian white{-}noise disorder. It was realized that $T$ exhibits giant mesoscopic fluctuations, while $\ln T$ is a self-averaging quantity: the variance of its distribution function is inversely proportional to the system's length. The transmission $T$ is not a self-averaging quantity, which is manifested in a small value of $\ln \Braket{T}/\Braket{\ln T}<1$. We find that the correlations in the pinned medium suppress the fluctuations. {Namely, c}orrelations reduce the variance of the $\ln T$ distribution function. {Moreover, t}he transmission $T$ remains to be a non-self-averaging quantity: the ratio $\ln \Braket{T}/\Braket{\ln T}\approx 0.90$ remains smaller, {\it but} quite close to1. We infer that the pinning-induced correlations suppress the difference between the optimal and typical disorder configurations, which determine $\ln \Braket{T}$ and $\Braket{\ln T}$, respectively. 

We find the transmission amplitude in a broad range of frequencies covering all the said regimes. We propose that this physics can be probed in long Josephson-junction arrays~\cite{Gurarie2004,Vogt2015,Cedergren2017,Bard2017} by measuring their microwave impedance~\cite{Houzet2019,Kuzmin2019}. {A combined effect of phase slips and disorder in these arrays may drive a transition into a ``glassy'' insulating state, which retains short-range superfluid correlations~\cite{Giamarchi1988}, and hence called Bose glass~\cite{Fisher1989}. In the classical limit of a high impedance array, which is the focus of our work, the static properties of the Bose glass lead to its universal dynamic response. Our theory facilitates the use of Josephson-junction arrays as a quantum simulation platform, allowing for the investigation of the Bose-glass phase and possibly observing the glass-superfluid transition taking place at smaller impedance.}

The rest of this article is organized as follows. In Sec.~\ref{sec:model}, we {characterize the static correlation of the pinned medium and we} define the plasmon scattering problem within the random phase sine-Gordon model. In Sec.~\ref{sec:semicalss}, we derive an analytical formula for the transmission in the regime {of weak disorder (large frequency)} using the Fokker-Planck method; we compare it with the numerical result. In Sec.~\ref{sec:localized}, we discuss our numerical results on the statistics of the transmission in the strongly localized regime {(small frequency)}. Finally, our results are summarized in Sec.~\ref{sec:discussion}.

\section{Model}
\label{sec:model}

In this section, we introduce the 1D random phase sine-Gordon model using, as an example, a Josephson-junction array connected to waveguides at its ends. 
Then we formulate the scattering problem for plasmons propagating through the array in the classical limit.
The numerics is used to find the average and the second moment of the effective disorder potential that appears in the linear wave equation for plasmons.
\subsection{Lagrangian}

The Lagrangian that describes the setup of Fig.~\ref{fig:setup} is a 1D random phase sine-Gordon model, which consists of two terms,
\begin{align}
\label{eq:L}
    {\cal L}={\cal L}_\mathrm{free}+{\cal L}_\mathrm{int}.
\end{align}
The harmonic term consists of the kinetic and elastic contributions,
\begin{align}
\label{eq:L_harm}
    {\cal L}_\mathrm{free}
    =\int dx~\frac \hbar{2\pi K(x)}
    \left[\frac{1}{v(x)}\dot\theta^2-v(x)(\partial_x\theta)^2\right],
\end{align}
where $\theta(x,t)$ is a local field associated with the accumulated charge, $\partial_x\theta$ is the one-dimensional charge density. It describes the propagation of waves in an inhomogeneous medium with local plasmon velocity $v(x)$ and local admittance $(4e^2/\pi \hbar)\cdot K(x)$.
In the waveguides ($x<0$ or $x>d$), $v(x)=v_0$ and $K(x)=K_0$;
in the medium ($0<x<d$), $v(x)=v$ and $K(x)=K$. Here $d$ is the length of the Josephson-junction array.
The harmonic term can be used to describe plasmons with wavelengths exceeding the screening length, $\ell_{\rm sc}=a\sqrt{C/C_g}$, where $a$ is the unit cell length in the array, and $C$ and $C_g$ are the junction and ground capacitances along the array (typically $\ell_{\rm sc}\gg a$).
Furthermore, the low-frequency impedance of the array is given by $K=\pi\sqrt{E_J/8E_g}$ with $E_g=e^2/2C_g$, and the plasmon velocity is $v=a\sqrt{8E_JE_g}/\hbar$.
Using the continuity of $\theta$ and $(v/K)\partial_x\theta$ at an interface between the waveguides and the medium, we find that the probabilities of the plasmon wave reflection and transmission at that interface are determined by the impedance mismatch, 
\begin{align}
\label{eq:R0T0}
    R_0=\left(\frac{K-K_0}{K+K_0}\right)^2,\quad T_0=\frac{4KK_0}{(K+K_0)^2}.
\end{align}
The interaction term in Eq.~\eqref{eq:L}, 
\begin{align}
    \label{eq:L_int}
    {\cal L}_\mathrm{int}
    =\int_0^ddx~\Lambda\cos(2\theta+\chi),
\end{align}
describes the pinning of the charge density. At $E_J\gg E_c$ 
\begin{align}
\Lambda=\frac{8E_c}{\sqrt{\pi}a} \left(\frac{2 E_J}{ E_c}\right)^{3/4}e^{-\sqrt{ 8{ E_J}/{ E_c}}},
\end{align}
where $E_c=e^2/2C$ is the charging energy of a junction. The random background charge introduces phase $\chi$ in the phase-slip amplitude~\cite{Ivanov2001,Friedman2002} due to the Aharonov-Casher effect; $\chi$ is a random variable with a short range of correlations.
In the experiment, the timescale for the variations of the configuration of background charges largely exceeds the relevant time scale for plasmon propagation. Therefore we will assume the configuration to be static in all subsequent calculations. On the other hand, the averaging time of an experiment may be sufficiently long for the disorder averaging over different configurations to be effectively realized on that timescale~\cite{Gurarie2004,Houzet2019}.

The characteristic Larkin length in the medium and the corresponding energy scale are determined~\cite{Larkin1970,Giamarchi2009,Houzet2019} by the competition between the elastic and disorder terms in Eqs.~\eqref{eq:L}, \eqref{eq:L_harm}, and \eqref{eq:L_int},
\begin{align}
\label{xi}
    \xi=\left(\frac{\hbar v}{2\pi K\Lambda\sigma}\right)^{2/(3-2K)}\quad\mathrm{and}\quad \Omega=\frac v\xi,  
\end{align}
respectively. Here we assumed the random background charges to be characterized by vanishing averages, $\Braket{\cos\chi(x)}=\Braket{\sin\chi(x)}=0$, and dispersion
\begin{align}
\label{eq:disorder}
\sigma^2=\int dx\Braket{\cos\chi(x)\cos\chi(0)}=\int dx\Braket{\sin\chi(x)\sin \chi(0)}.
\end{align}
{The collective pinning regime corresponds to the condition $\xi\gg\sigma^2$.
The limit $K\ll 1$ in Eq.~\eqref{xi} corresponds to the classical pinning. In this limit, one may crossover from the regime of the collective pinning to that of individual pinning by increasing $\Lambda$. A finite $K$ allows for quantum fluctuations~\cite{Suzumura1983,Houzet2019} favoring longer $\xi$ at the same value of $\Lambda$. The divergence of $\xi$ at $K=3/2$ marks the transition} between the Bose glass and the superfluid phase~\cite{Giamarchi1988}.

\subsection{Scattering problem in the classical limit}

To formulate the linear scattering problem in the classical limit ($K\to0$ at a fixed $K/\hbar$){, i.e., deep in the Bose-glass phase,} we represent the field $\theta(x,t)$ as the sum of a static part, $\bar\theta(x)$, plus small oscillations around it,
\begin{align}
    \theta(x,t)=\bar\theta(x)+\psi(x)e^{-i\omega t}.
\end{align}
The static field $\bar\theta$ minimizes the energy functional
\begin{align}
\label{eq:E}
    {\cal E}[\theta]
    =\int_0^d dx~\left[\frac {\hbar v}{2\pi K}(\partial_x\theta)^2-\Lambda\cos(2\theta+\chi)\right],
\end{align}
with boundary conditions $\partial_x\theta(0)=\partial_x\theta(d)=0$ for the charge density. The oscillatory component, $\psi(x)$, takes the asymptotic form of a scattering state at frequency $\omega$ in the waveguides,
\begin{align}
    \label{eq:small_oscillation}
    \psi(x)
    =\left\{\begin{array}{ll}
    e^{i\omega x/v_0}+r(\omega)e^{-i\omega x/v_0}, & x<0, \\
    t(\omega)e^{i\omega (x-d)/v_0}, & x>d.
    \end{array}\right.
\end{align}
Here $r(\omega)$ and $t(\omega)$ are the (elastic) reflection and transmission amplitudes, respectively.
The wavefunction $\psi(x)$ solves the wave (Schr\"odinger-like) equation{~\cite{footnote-disorder}}
\begin{align}
    \label{eq:schroedinger}
    \omega^2\psi(x)=-v^2\partial_x^2\psi(x)+{\cal V}(x)\psi(x),
\end{align}
with the effective disorder potential
\begin{align}
\label{eq:potential}
    {\cal V}(x)=(4\pi Kv\Lambda/\hbar) \cos\left[2\bar\theta(x)+\chi(x)\right]
\end{align}
in the medium, together with boundary conditions
\begin{subequations}
\label{eq:BC}
\begin{align}
    1+r(\omega)&=\psi(0), \label{eq:BC1}\\
    i\omega{K}/{K_0}(1-r(\omega))&=v\partial_x\psi(0), \label{eq:BC2}\\
    t(\omega)&=\psi(d), \\
    i\omega{K}/{K_0}t(\omega)&=v\partial_x\psi(d),
\end{align}
\end{subequations}
which result from matching the solution $\psi(x)$ at $x=0$ and $x=d$, cf.~Eq.~\eqref{eq:small_oscillation}.

The scattering properties obtained by the solution of Eqs.~\eqref{eq:small_oscillation}-\eqref{eq:potential}, with $\bar\theta$ defined by minimization of Eq.~\eqref{eq:E}, depend on the frequency, medium's length, and impedance mismatch. 
Remarkably, {we found numerically {(see Appendix~\ref{app:numerics} for details)} that} the results {\it{at any frequency $\omega$}} depend on the disorder only through the dispersion $\sigma$ defined in Eq.~\eqref{eq:disorder}, under the assumption of short-range correlations (this finding is in correspondence with the central limit theorem). 
Thus the results for the transmission and reflection amplitudes are universal once the length $d$ and frequency $\omega$ are expressed in units of $\xi$ and $\Omega=v/\xi$, see Eq.~\eqref{xi}.

\subsection{Effective disorder potential}

\begin{figure}
\includegraphics[width=0.9\columnwidth]{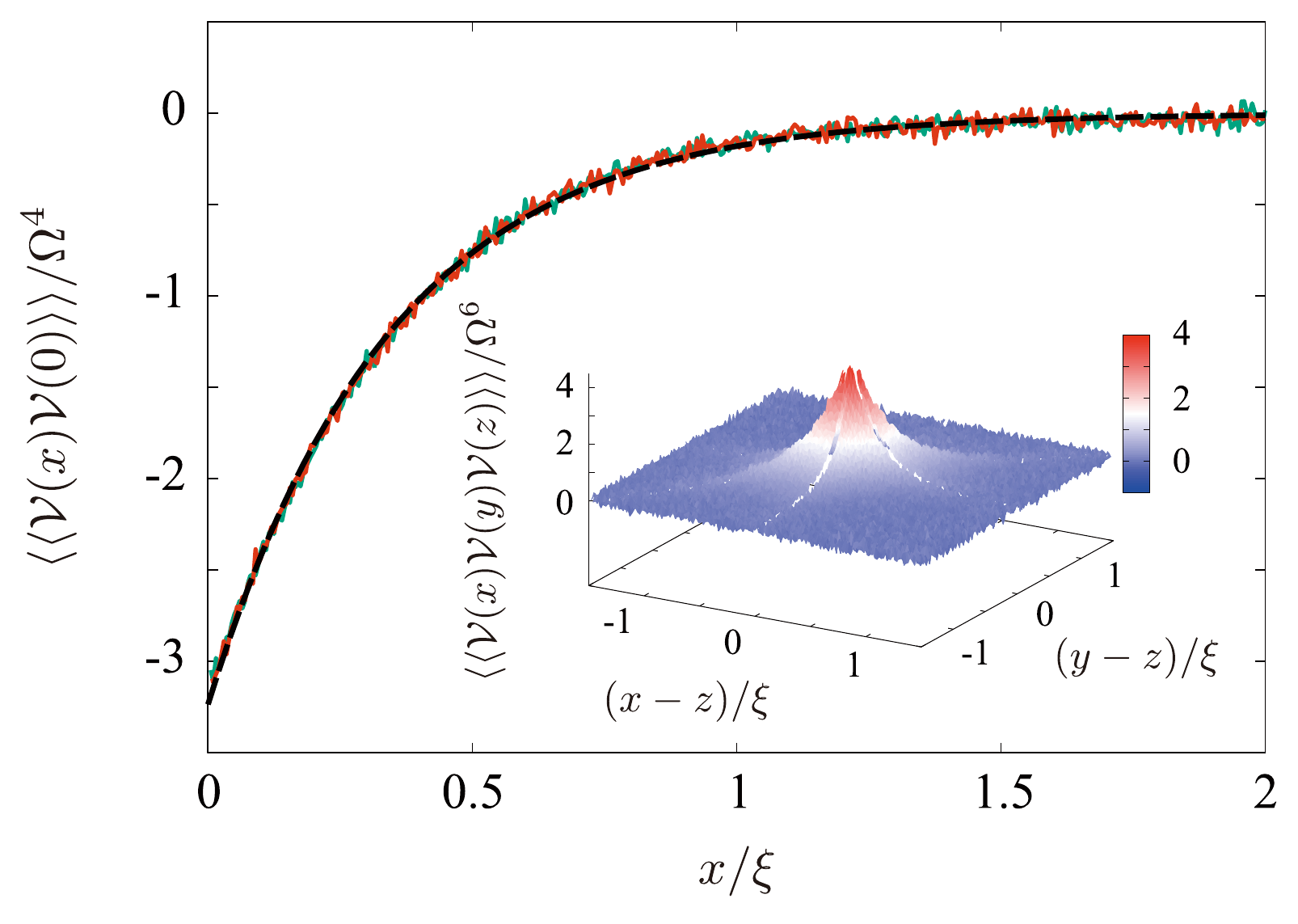}
\caption{\label{fig:potential}
{The {smooth part of the} second-order correlation function of the effective disorder potential} at $d/\xi=20$.
The green and red lines are the numerically obtained curves for the random ``phase'' and ``box'' models (see Appendix~\ref{app:numerics}), respectively; they cannot be distinguished from each other. The dashed lines represent the fitting functions $-w(x)=-ce^{-|x|/a}$ ($c=3.24,~a=0.346$). {The inset shows the third-order cumulant at non-coinciding points.}}
\end{figure}

The effective potential \eqref{eq:potential} depends on the solution $\bar \theta(x)$ of a nontrivial optimization problem defined by the functional \eqref{eq:E}. It is thus expected to have a complex pattern of non-Gaussian correlations of which, to the best of our knowledge, little is known. The reality condition of the eigenspectrum of the wave equation \eqref{eq:schroedinger} means that the local potential has a well defined sign and, hence, a positive mean value, $\langle {\cal V}(x)\rangle>0$. The potential $\cal{V}$ inherits a finite correlation length $\xi$ from $\overline\theta$, despite the underlying disorder $\chi(x)$ being a short-ranged one. We characterize the {spatial} correlations in ${\cal V}(x)$ by 
{considering its second cumulant and expressing it in the form}
\begin{align}
    \label{eq:correlator}
    \Braket{\Braket{{\cal V}(x){\cal V}(y)}}
    &\equiv\Braket{\left({\cal V}(x)-\Braket{{\cal V}}\right)\left({\cal V}(y)-\Braket{{\cal V}}\right)} \\
    &= \Omega^4\left[4\delta\left(\frac{x-y}{\xi}\right)-w\left(\frac{x-y}{\xi}\right)\right]. \nonumber
\end{align}
{Here, the introduced function $w(x)$ is smooth and decays at large scales. W}e find $w(x/\xi)$ numerically, see Fig.~\ref{fig:potential} and Appendix~\ref{app:numerics}. {We note that the random potential ${\cal V}(x)$ is non-Gaussian; its third-order cumulant is shown in the inset of Fig.~\ref{fig:potential}.} The Fourier component of $\Braket{\Braket{{\cal V}(x){\cal V}(0)}}$ at small momentum (on the scale of $1/\xi$) is reduced by a factor $1-\eta$ with {$\eta=1/4\int dy w(y)\approx 0.561$} compared to its large-momentum value, which is determined by the first term in the r.h.s.~of Eq.~\eqref{eq:correlator}. 
The property $\eta\neq 0$ is the consequence of the collective pinning mechanism. In the large-frequency range of interest in Sec.~\ref{sec:semicalss}, $\eta$ is the only parameter needed to predict the statistics of the scattering amplitudes. 
{On the other hand, we illustrate in Sec.~\ref{sec:localized} that the signature properties of the reflection and transmission amplitudes at low frequencies cannot be accounted for by a colored Gaussian disorder that would have the same average and second moment as the effective disorder potential. Therefore we attribute these properties to the non-Gaussianity of the potential ${\cal V}(x)$ induced by pinning.}

\section{Weakly localized regime (high frequency)}
\label{sec:semicalss}

In this section, we derive an analytical formula for the transmission probability {in the regime of high frequency, in which disorder can be treated perturbatively;} we  also compare it with the results of numerical calculations. 

\subsection{Mapping to the Dirac equation}
\label{sec:dirac}

We first show that, at large frequency, the wave equation~\eqref{eq:schroedinger} is equivalent to a Dirac equation in Gaussian random fields. For this, we introduce two functions, $R(x)$ and $L(x)$, such that
\begin{subequations}
\begin{align}
    \psi(x)&=R(x)+L(x), \\
    \partial_x\psi(x)&=i\frac{\omega}{v}[R(x)-L(x)].
\end{align}
\end{subequations}
Using $\partial_x\psi(x)=\partial_xR(x)+\partial_xL(x)$ and Eq.~\eqref{eq:schroedinger}, we deduce
\begin{subequations}
\label{eq:RL}
\begin{align}
    \omega R(x)&=-iv\partial_xR(x)+\frac{{\cal V}(x)}{2\omega}[R(x)+L(x)],\\
    \omega L(x)&=iv\partial_xL(x)+\frac{{\cal V}(x)}{2\omega}[R(x)+L(x)].
\end{align}
\end{subequations}
Functions $R(x)$ and $L(x)$ have the meaning, respectively, of the right- and left-moving components of the wave function. The terms $\propto{\cal V}(x)$ in Eqs.~\eqref{eq:RL} result in scattering between the components. In the weak-disorder regime, only the forward- and back-scattering amplitudes evaluated in the Born approximation determine the scattering properties of the medium. At $\omega\gg\Omega$, these amplitudes are associated with the harmonics of the random potential near momenta $k = 0$ and $k =2\omega/v\gg 1/\xi$, respectively. That allows us to approximate
\begin{align}
    \label{eq:V_components}
    \frac{{\cal V}(x)}{2\omega}\approx\Delta_0(x)+\Delta(x)e^{2i\omega x/v}+\Delta^*(x)e^{-2i\omega x/v},
\end{align}
where $\Delta_0(x)$ and $\Delta(x)$ are slowly varying random functions (on the scale of $v/\omega$). We find that the slow components of Eq.~\eqref{eq:RL} yield a Dirac equation,
\begin{align}
    \label{eq:dirac}
    \left[-iv\partial_x\tau_3+\Delta_0(x)+\Delta_1(x)\tau_1+\Delta_2(x)\tau_2\right] \Phi(x)=0,
\end{align}
with $\Phi(x)=[R(x)e^{-i\omega (x-d)/v},L(x)e^{i\omega (x-d)/v}]^T$, $\Delta(x)=\Delta_1(x)+i\Delta_2(x)$ with real fields $\Delta_1(x)$ and $\Delta_2(x)$, and Pauli matrices $\tau_{1,2,3}$.
Using Eq.~\eqref{eq:correlator}, the Born approximation applied to scattering off a potential given by Eq.~\eqref{eq:V_components} yields the forward- and back-scattering lengths, 
\begin{align}
\label{eq:lengths}
\ell_0(\omega)=\ell_\pi(\omega)/(1-\eta)\quad\mathrm{and}\quad\ell_\pi(\omega)=\xi (\omega/\Omega)^2, \end{align}
respectively. Thus the correlation between the random background charge and the static charge $\propto \partial_x\bar\theta$ induced by it
tends to weaken the forward-scattering rate and introduces a (quantitative) difference between $\ell_0$ and $\ell_\pi$. As $\ell_0,\ell_\pi\gg\xi$ in the considered high-frequency regime, the random fields in Eq.~\eqref{eq:dirac} may be viewed as Gaussian ones, due to the sampling over a large number of correlated regions, each of which having a length scale $\sim\xi$. The needed two Fourier components of the correlation function of $\cal V$ are reproduced by the Fourier components of the respective correlation functions
\begin{subequations}
\label{eq:correlator-Dirac}
\begin{align}
    \Braket{\Delta_0(x)\Delta_0(y)}&=\frac{\Omega^4}{\omega^2}(1-\eta)\delta\left(\frac{x-y}{\xi}\right), \\
    \Braket{\Delta(x)\Delta^*(y)}&=\frac{\Omega^4}{\omega^2}\delta\left(\frac{x-y}{\xi}\right).
\end{align}
\end{subequations}

{Both fields $\Delta_0$ and $\Delta$ induce a random phase between right- and left-movers. The contribution from $\Delta_0$ to the phase acquired over a distance $x$ is readily obtained, $(2/v)\int_0^x dy \Delta_0(y)$; its variance grows as $4x/\ell_0$. As we will see below, the additional contribution from $\Delta$ is $2x/\ell_\pi$; it only quantitatively modifies the result. Phase scrambling occurs in a medium of length $d\gg\ell_0,\ell_\pi$. Using Eq.~\eqref{eq:lengths}, this condition is equivalent to a small frequency condition, $\omega\ll\omega_{\rm cr}$ with crossover frequency $\omega_{\rm cr}=\Omega\sqrt{d/\xi}\gg\Omega$. Reference~\cite{Houzet2019} addressed the role of $\omega_{\rm cr}$ in the visibility of oscillations in the reflection amplitude. Below we find a similar effect in the frequency dependence of the transmission coefficient, see Sec.~\ref{sec:Ttun}.}

{In addition,} $\Delta$ is responsible for the plasmons localization, due to the waves' back-scattering it generates.

\subsection{Fokker-Planck formalism}
\label{sec:FP}

To proceed further, we rely on the Fokker-Planck formalism that was developed to predict the statistics of scattering properties of waves subject to Gaussian white{-}noise disorder. The previously known results~\cite{Gertsenshtein1959,Lifshitz1988,Abrikosov1976,Antsygina1981,beenakker1993} assume a perfect impedance matching. Below we generalize them to the case of a finite impedance mismatch.

For a scattering state, the wave functions in the waveguides and the medium are related through 
\begin{align}
   \label{eq:barrier_scattering}
    \left(\begin{array}{c}
    R(0) \\
    r(\omega)
    \end{array}\right)
    = S_0
    \left(
    \begin{array}{cc}
    1 \\
    L(0)
    \end{array}
    \right), \quad
    \left(\begin{array}{c}
    t(\omega) \\
    L(d)
    \end{array}\right)
    = S_0^T
        \left(
    \begin{array}{cc}
    R(d) \\
    0
    \end{array}
    \right).
\end{align}
Here a phase factor $e^{-i\omega d/v}$ was absorbed in $t(\omega)$ and we introduced the scattering matrix at each interface,
\begin{align}
    S_0
        = \left(
    \begin{array}{cc}
    t_0 & r_0 \\
    -r_0 & t_0
    \end{array}
    \right)
\end{align}
with $r_0=\sqrt{R_0}$ and $t_0=\sqrt{T_0}$, see Eq.~\eqref{eq:R0T0}~{\cite{footnote-refl0}}. 
This allows us to relate the scattering amplitudes with  
the Ricatti variable, 
\begin{align}
z(x)=[L(x)/R(x)]e^{2i\omega (x-d)/v}, 
\end{align}
at the ends of the medium,
\begin{equation}
\label{eq:r}
r(\omega)=\frac{-r_0+z(0)e^{2i\omega d/v}}{1-r_0z(0)e^{2i\omega d/v}}\quad{\rm with}\quad z(d)=r_0.
\end{equation}
Interestingly, the Ricatti variables that appear in Eq.~\eqref{eq:r} are related through a first-order nonlinear stochastic differential equation derived from Eq.~\eqref{eq:dirac},
\begin{equation}
\label{eq:ricatti}
-iv\partial_x z(x)=2\Delta_0(x) z(x)+\Delta^*(x)+\Delta(x) z^2(x).
\end{equation}
Thus $r(\omega)$ and, subsequently, the transmission probability $T(\omega)=1-|r(\omega)|^2$ are expressed in terms of a solution of Eq.~\eqref{eq:ricatti} with a given boundary condition.

Taking advantage of the Gaussian correlators \eqref{eq:correlator-Dirac}, we derive the Fokker-Planck equation for the conditional distribution probability $P(x,\theta_1,\theta_2)$, where we introduced the decomposition of the Ricatti variable into its amplitude and phase, $z=e^{i\theta_1}e^{-\theta_2}$ {($\theta_2\geq 0$, such that $|r(\omega)|\leq 1$~\cite{Lifshitz1988})}. It reads 
\begin{align}
    \label{eq:FP}
    -\frac{\partial P}{\partial x}
    =
        \frac{2}{\ell}\frac{\partial^2P}{\partial\theta_1^2}
    +\frac{1}{\ell_\pi}\frac{\partial^2}{\partial\theta_2^2}\left(\sinh^2\theta_2P\right),\quad 0<x<d,
\end{align}
with
\begin{align}
\label{eq:mfp}
\ell(\omega)=\left(\frac 1{\ell_0(\omega)}+\frac 1{2\ell_\pi(\omega)}\right)^{-1}=\frac{2}{3-2\eta}\ell_\pi(\omega)
\end{align}
(see Appendix~\ref{App:FP} for the derivation). The solution of Eq.~\eqref{eq:FP} is separable, $P(x,\theta_1,\theta_2)=P_1(x,\theta_1)P_2(x,\theta_2)$; each factor satisfies
\begin{subequations}
\begin{align}
    \label{eq:FP_P1}
   - \frac{\partial P_1}{\partial x}
    &=
    \frac2\ell\frac{\partial^2P_1}{\partial\theta_1^2}, \\
    \label{eq:FP_P2}
    -\frac{\partial P_2}{\partial x}
    &=\frac 1{\ell_\pi}\frac{\partial^2}{\partial\theta_2^2}\left(\sinh^2\theta_2P_2\right),
\end{align}    
\end{subequations}
with initial conditions $P_1(d,\theta_1)=\delta(\theta_1)$ and $P_2(d,\theta_2)=\delta(\theta_2-\theta_0)$ with $\theta_0=-\ln r_0$.
Noteworthy, the statistics of $\theta_2$ is only sensitive to back-scattering, thus reflecting the medium's localization properties; it is insensitive to Larkin's physics. By contrast, the statistics of $\theta_1$, {which describes the random phase between right- and left-movers,} depends both on forward- and back-scattering; as $\eta\neq 0$, it is sensitive to Larkin's physics.

Equation~\eqref{eq:FP_P1} is a standard diffusion equation, its solution at $x=0$ is
\begin{align}
\label{eq:FP_P1-sol}
 P_1(\theta_1)\equiv  P_1(0,\theta_1)=\frac{1}{\sqrt{8 \pi d/\ell}} \exp\left(-\frac{\theta_1^2}{8d/\ell}\right);
\end{align}
the variance $\Braket{\theta^2_1}=4d/\ell_0+2d/\ell_\pi$ includes the contribution from $\Delta_0$ discussed in Sec.~\ref{sec:dirac}, plus the contribution from $\Delta$, which has the same order of magnitude. The solution of Eq.~\eqref{eq:FP_P2} can also be obtained, 
\begin{align}
\label{eq:P2}
P_2(\theta_2)&\equiv P_2(0,\theta_2)\\
&=\frac{e^{-d/4\ell_\pi}}{4\sinh^2\theta_2}
\int_0^\infty dkk\tanh\frac{\pi k}2
P_{-\frac12+i\frac k2}(\coth \theta_2)\nonumber \\
&\qquad\qquad\qquad\qquad\times P_{-\frac12+i\frac k2}(\coth \theta_0)e^{-{k^2d/4\ell_\pi}},
\nonumber 
\end{align}
where $P_{-\frac12+i\frac k2}$ is the Legendre function of the first kind (see Appendix~\ref{App:P2} for the derivation).
The ensemble-averaged transmission is then given as
\begin{align}
\label{eq:Tav}
    \Braket{T}
    = \int d\theta_1\int d\theta_2~
    P_1(\theta_1)P_2(\theta_2)T(\theta_1,\theta_2),
\end{align}
where 
\begin{align}   
\label{eq:T12} 
    T(\theta_1,\theta_2)&=T_0\frac{1-e^{-2\theta_2}}{1+r_0^2e^{-2\theta_2}-2r_0\sin(2\omega d/v+\theta_1)e^{-\theta_2}}
\end{align}
is obtained by inserting $z(0)=e^{i\theta_1-\theta_2}$ into Eq.~\eqref{eq:r} and using $T=1-|r|^2$. Using the above equations, below we obtain simpler formulas in various cases.

\subsubsection{Perfect impedance matching, $T_0=1$}
{At $T_0=1$, Eq.~\eqref{eq:T12} simplifies, $T(\theta_1,\theta_2)=1-e^{-2\theta_2}$, such that the statistics of $\theta_2$ fully determines the transmission coefficient.} We use $P_\nu(1)=1$ and $\int _1^\infty dx{P_{-1/2+ik/2}(x)}/{(1+x)}={\pi}/{\cosh(k\pi/2)}$ 
(Eq.~7.131.1 in Ref.~\cite{Gradshteyn}) to find
\begin{align}   
\label{eq:Tbal}
    \Braket{T}=\int_0^\infty dk \frac{\pi k}2\frac{\tanh({\pi k}/2)}{\cosh(\pi k/2)}e^{-(1+k^2)d/4\ell_\pi}.
\end{align}
At $d\ll \ell_\pi$, $\Braket{T}\approx1$ as expected for a ballistic junction. At $d\gg\ell_\pi$, the ensemble-averaged transmission coefficient is
\begin{align}
\label{eq:Gertshenstein}
  \Braket{T}\approx   \frac {\pi^{5/2}}{2}\left(\frac{\ell_\pi}d\right)^{3/2}e^{-d/4\ell_\pi},
\end{align}
in agreement with Ref.~\cite{Gertsenshtein1959}. This results was rediscovered in {Refs.~\cite{Abrikosov1976,Antsygina1981}}. The multichannel generalization was performed in Ref.~\cite{beenakker1993}. Notably, the frequency dependence of $\Braket{T}$ is smooth (no oscillation) and fully captured by $\ell_\pi(\omega)$.

\subsubsection{Asymptote at $d\gg\ell(\omega)$ and arbitrary $T_0$}

The sine term in Eq.~\eqref{eq:T12} produces Fabry-P\'erot oscillations of the transmission, with period $\pi v/d$, provided that the impedance mismatch is finite. At $d\gg \ell(\omega)$, the dispersion of $\theta_1$ given by Eq.~\eqref{eq:FP_P1-sol} washes out the oscillations on average. In that regime, $\theta_1$ can be taken as uniformly distributed, and Eq.~\eqref{eq:Tav} simplifies to
\begin{align}
\label{eq:Tavbal}
    \Braket{T}
    = \int d\theta_2 P_2(\theta_2)T(\theta_2),\quad T(\theta_2)=\frac{T_0(1-e^{-2\theta_2})}{1-R_0e^{-2\theta_2}}.
\end{align}
At $T_0=1$, the result reproduces Eq.~\eqref{eq:Tbal}.

\begin{figure}
\includegraphics[width=0.9\columnwidth]{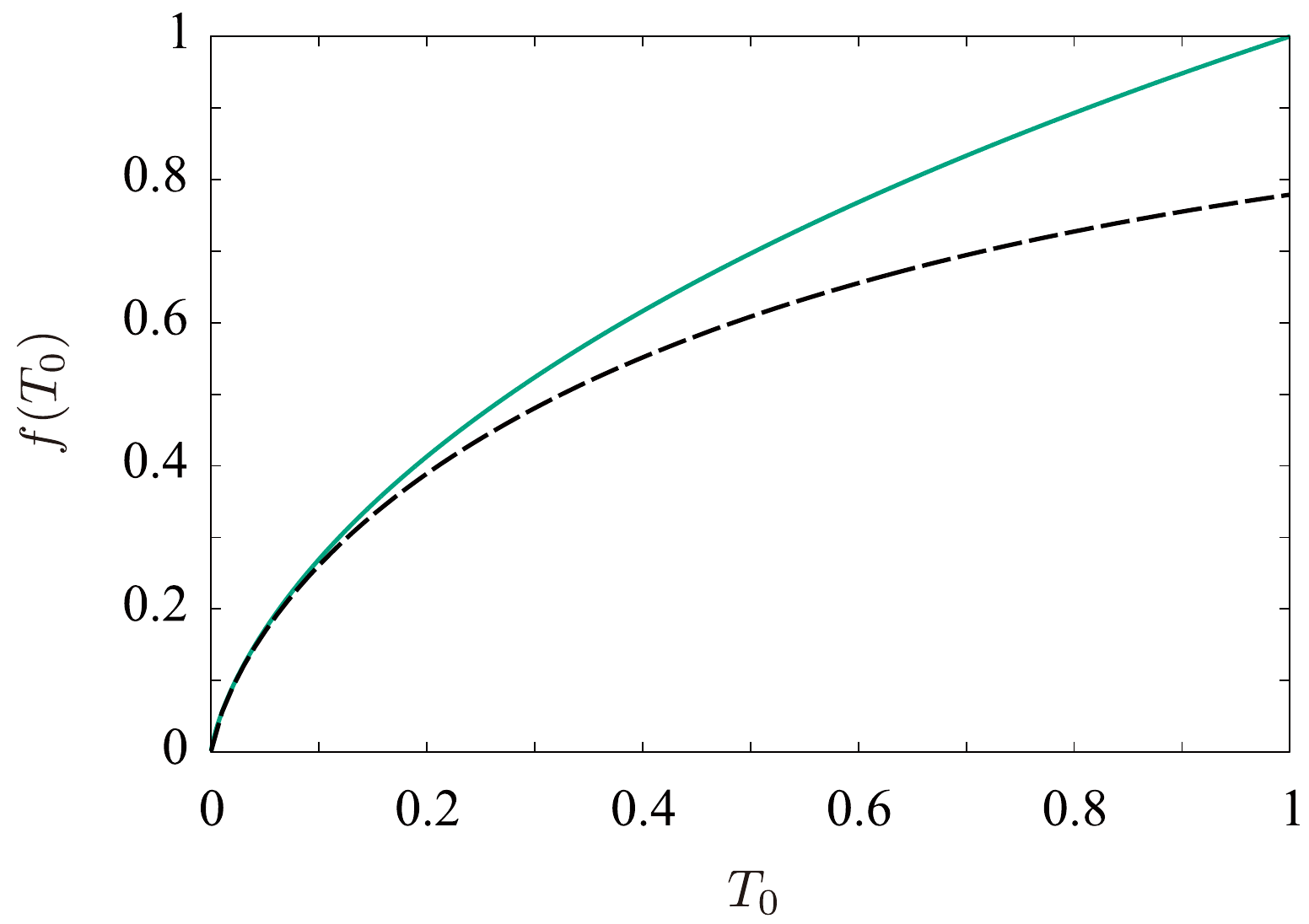}
\caption{\label{fig:fT0}
Graph of $f(T_0)=(4T_0/\pi^2)\mathbb{K}^2(1-T_0)$ (plain line) and its asymptote, Eq.~\eqref{eq:asymptote-fT0}, at $T_0\ll 1$ (dashed line).}
\end{figure}

Returning to an arbitrary $T_0$, we use Eq.~\eqref{eq:Tavbal} with $P_2(\theta_2)$ of Eq.~\eqref{eq:P2}. At $d\gg\ell_\pi$, the $k$-integral in Eq.~\eqref{eq:P2} is dominated by the region $k\ll 1$, so that we can replace 
\begin{equation}
P_{-1/2+ik/2}(x)\approx P_{-1/2}(x)=\frac {2\sqrt{2}}{\pi\sqrt{x{+}1} }\mathbb{K}\left(\frac{x-1}{x+1}\right)
\end{equation}
with complete elliptic integral $\mathbb{K}(m)=\int_0^{\pi/2} {d\phi}/{\sqrt{1-m\sin^2\phi}}$, and perform the $k$-integral explicitly. Then, the dependences on the length and the barriers' transmission decouple, and we get
\begin{equation}
\Braket{T}=\frac{\pi^{5/2}}{2}\left(\frac{\ell_\pi}d\right)^{3/2}e^{-d/4\ell_\pi}f(T_0)
\label{eq:transmission}
\end{equation}
with $f(T_0)=(4T_0/\pi^2)\mathbb{K}^2(1-T_0)$, see Fig.~\ref{fig:fT0}. In particular, $f(T_0=1)=1$, reproducing Eq.~\eqref{eq:Gertshenstein}, while 
\begin{equation}
\label{eq:asymptote-fT0}
f(T_0\ll 1)\approx\frac{T_0}{\pi^2}\ln^2\left(\frac{16}{T_0}\right).
\end{equation}
Note that, as in Eq.~\eqref{eq:Gertshenstein},  the $\omega$-dependence of the pre-exponential factor in Eq.~\eqref{eq:transmission} comes only via $\ell_\pi$. {The $\omega$-dependence of the transmission is again smooth in this regime.}

\subsubsection{Asymptote at $d\ll\ell(\omega)$ and $T_0\ll 1$}
\label{sec:Ttun}

At finite impedance mismatch and $d\ll\ell$, $T_0\neq 0$, Fabry-P\'erot oscillations do exist. If in addition the mismatch is large, $T_0\ll 1$, we may use the initial value for the $\theta_2$-distribution, $P_2(\theta_2)\approx\delta(\theta_2-T_0/2)$ to find
\begin{align}
\label{eq:T-tun}
\Braket{T}=\sum_n\int d\theta_1\frac{T_0^2}{T_0^2+\left[2({\omega-\omega_n})d/v+{\theta_1}\right]^2}P_1(\theta_1)
\end{align}
with $P_1$ of Eq.~\eqref{eq:FP_P1-sol}. Here $\omega_n=n\pi v/d$ are the frequencies at which the transmission has local maxima. Equation~\eqref{eq:T-tun} describes how, upon lowering $\omega$, the oscillations' Lorentzian lineshapes with half-width $T_0 v/2d$ (determined solely by the impedance mismatch) evolve into Gaussian lineshapes with half-width $v\sqrt{2\ln 2/d\ell}$ determined by the medium randomness. The crossover frequency, which is reached when $T_0\sim \sqrt{d/\ell(\omega)}$, is $\omega_{\rm cr}/T_0$. Remarkably, in the frequency range $\omega_{\rm cr}\ll\omega\ll\omega_{\rm cr}/T_0$ the width of the oscillations is sensitive to $\eta$, cf. Eq.~\eqref{eq:mfp}, and, henceforth, to the collective pinning.

\subsection{Discussion}
\label{sec:disc}

\begin{figure}
\flushleft{(a)}\\
\centering
\includegraphics[width=0.9\columnwidth]{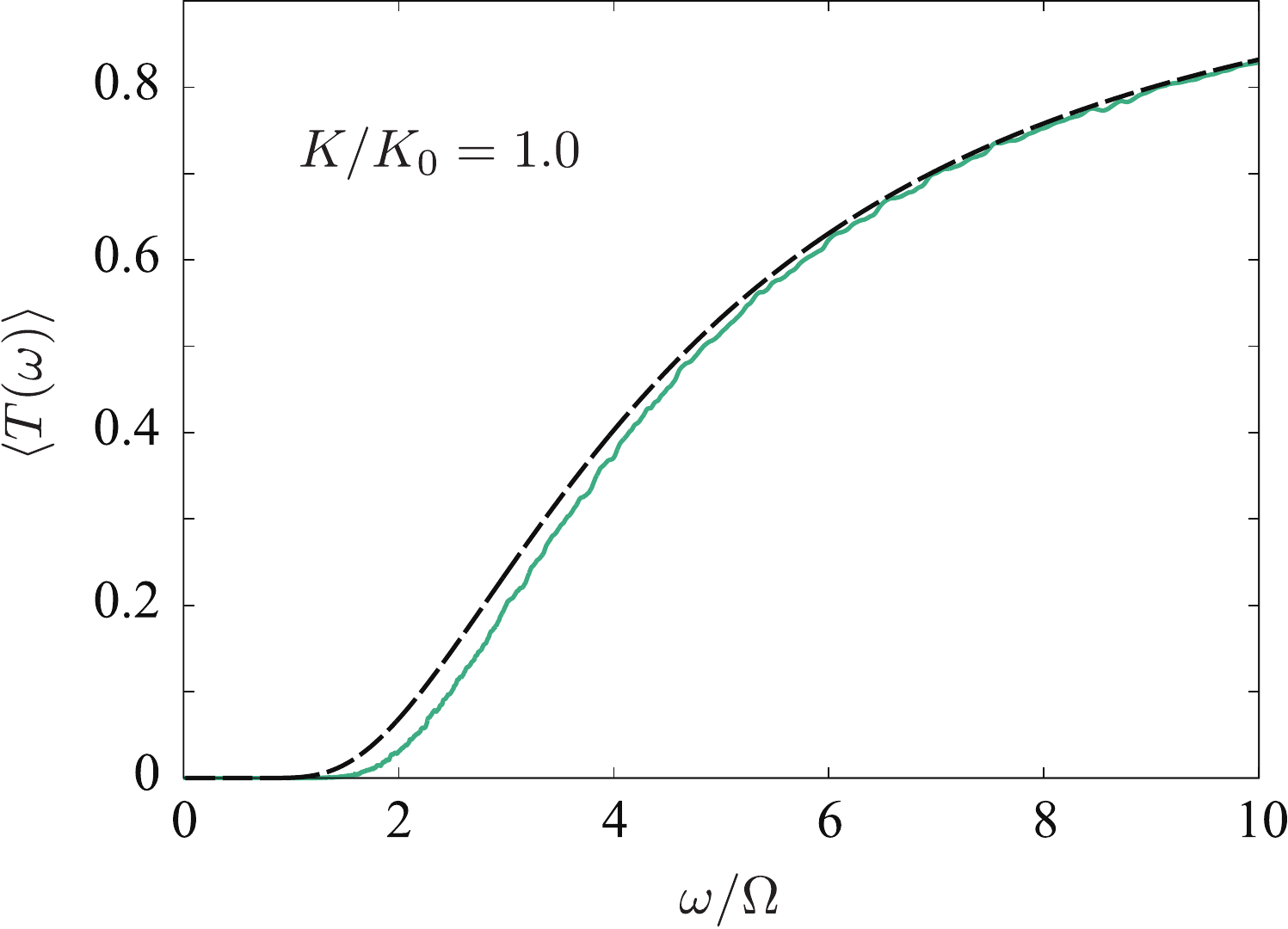}
\\
\flushleft{(b)}\\
\centering
\includegraphics[width=0.9\columnwidth]{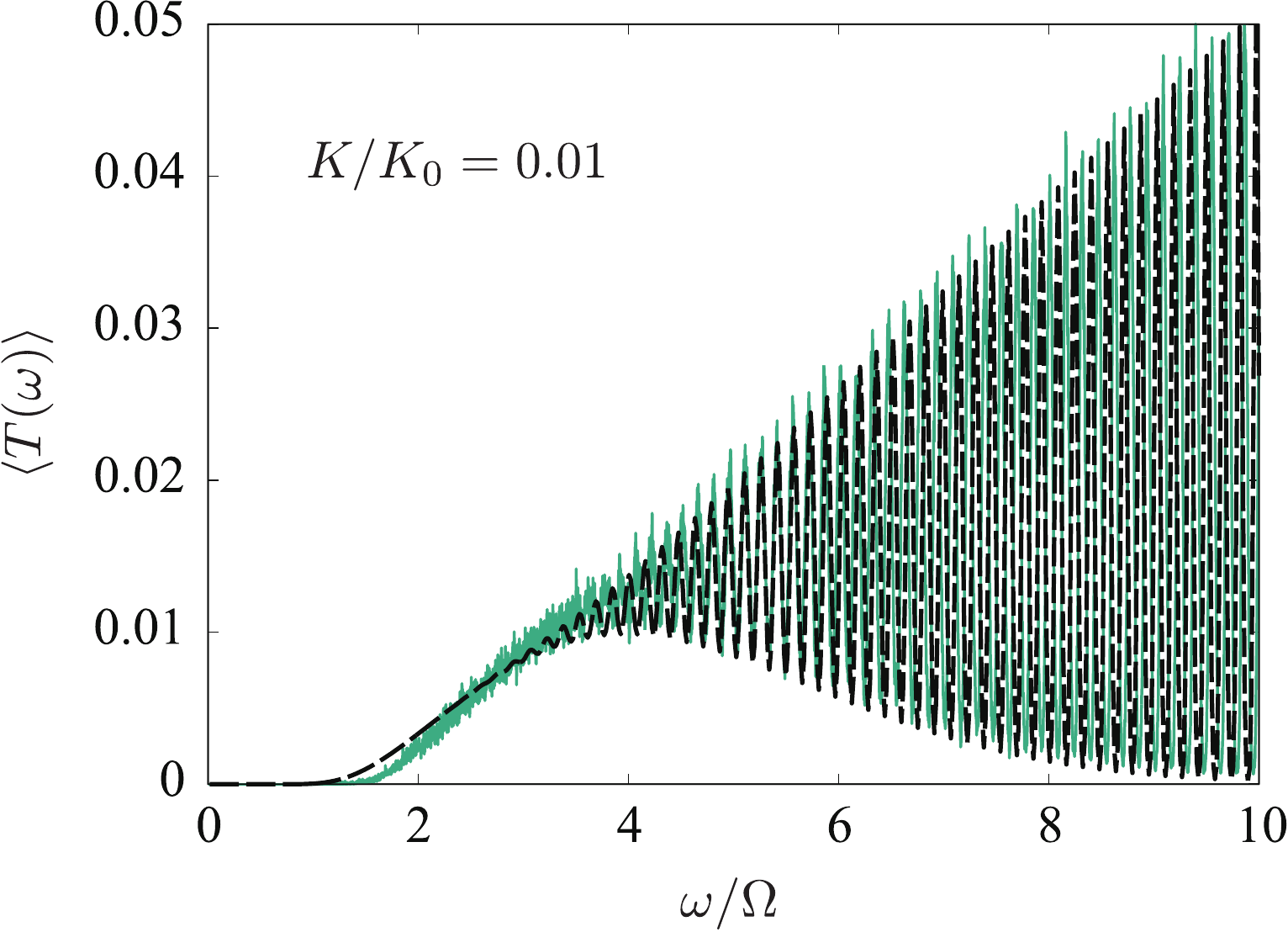}
\caption{\label{fig:T_semiclass}
Frequency dependence of $\Braket{T(\omega)}$ for (a) $K/K_0=1.0$ and (b) $K/K_0=0.01$ and $d/\xi=20$. The green dots are the numerics, the dashed line is Eq.~\eqref{eq:Tbal} in (a) and Eq.~\eqref{eq:Ttun-cross} in~(b).}\end{figure}

The brute-force numerics is compared with the predictions from the Fokker-Planck formalism in Fig.~\ref{fig:T_semiclass}. In that Figure, {Eq.~\eqref{eq:Tbal}} is used at perfect impedance matching, while we use
\begin{align}
\label{eq:Ttun-cross}
\Braket{T}=2\pi T_0 \sum_nP_1\left(\frac{2\omega_nd}v\right)\int d\theta_2\frac{2\theta_2}{2\theta_2+T_0}P_2(\theta_2),
\end{align}
valid in the frequency range $\Omega\ll \omega\ll \omega_{\rm cr}/T_0$, at $T_0\ll1$, see Figs. \ref{fig:T_semiclass}(a) and \ref{fig:T_semiclass}(b) respectively.

\begin{figure}
\flushleft{(a)}\\
\centering
\includegraphics[width=0.9\columnwidth]{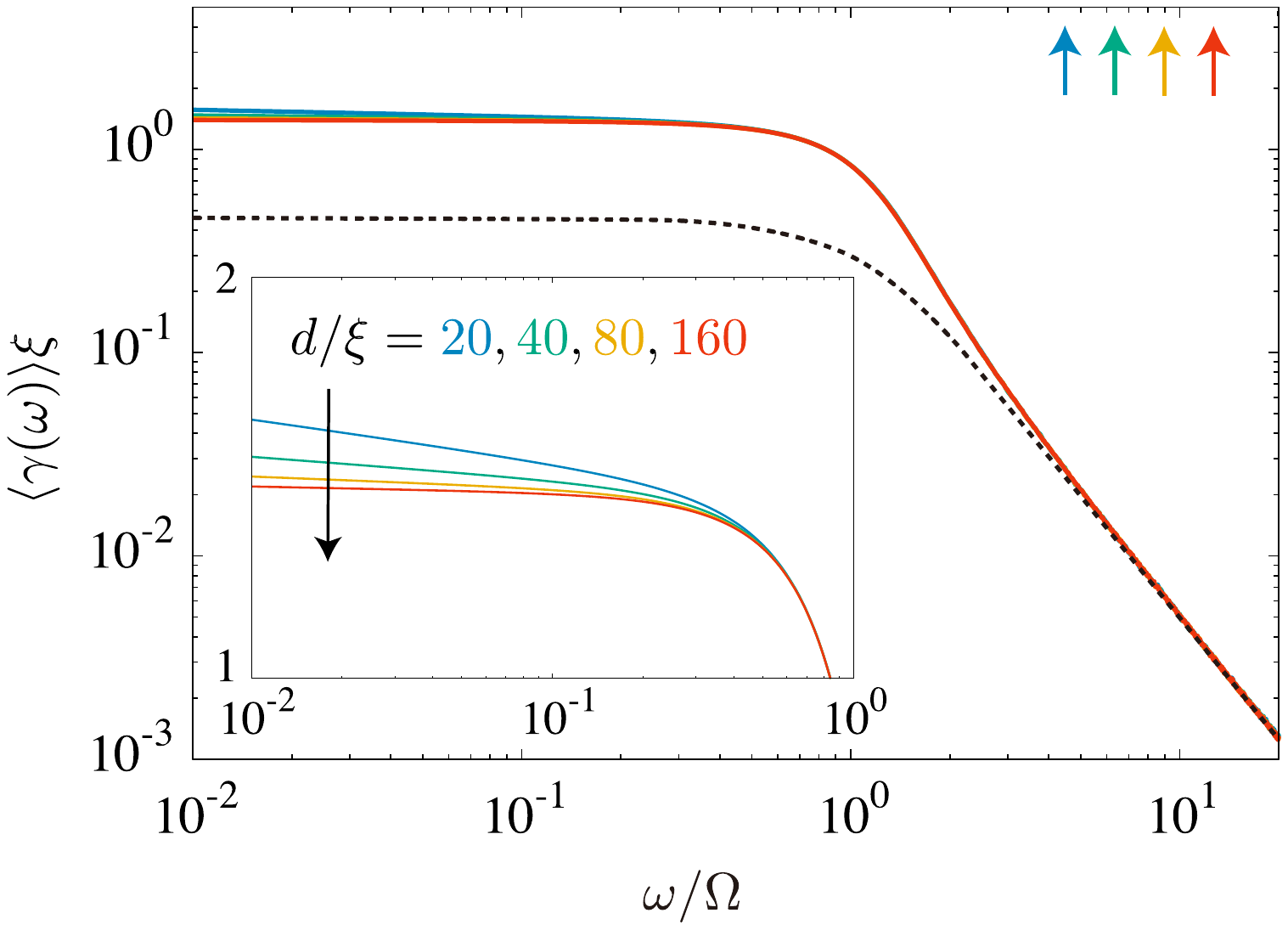}
\\
\flushleft{(b)}\\
\centering
\includegraphics[width=0.9\columnwidth]{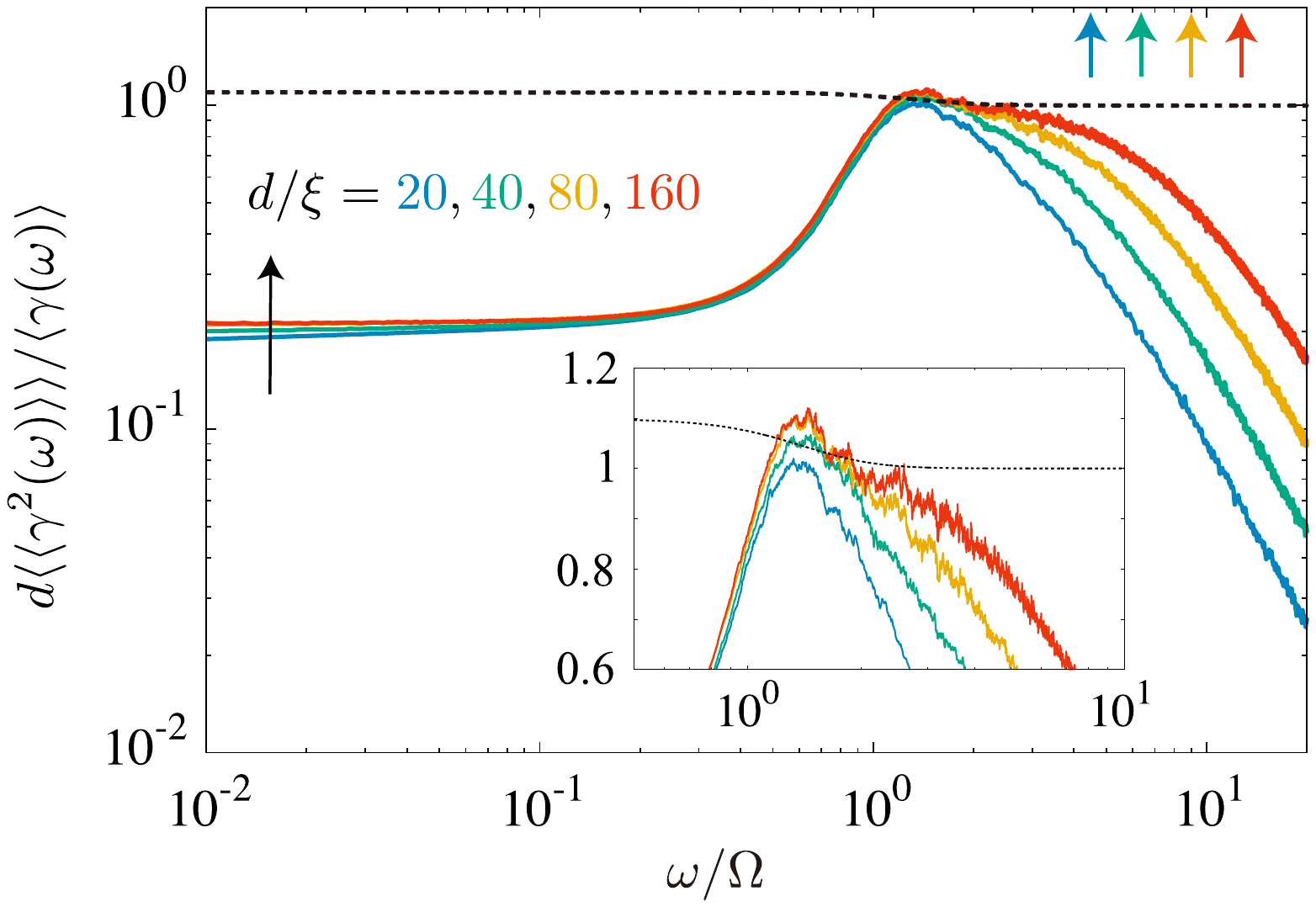}
\caption{\label{fig:lyapunov}
(a) The average of the Lyapunov exponent and (b) its variance as a function of frequency for different lengths $d/\xi=20,40,80,160$ and $K/K_0=1.0$. The frequency range covers the regimes of weak and strong plasmon localization, $\omega\gg\Omega$ and $\omega\ll\Omega$, respectively.
{Arrows indicate the crossover frequency to the ballistic regime, $\omega_{\rm cr}=\Omega\sqrt{d/\xi}$, for each length.} 
The inset is an enlarged view of the low (a) and intermediate (b) frequency regions.
{The dotted lines in (a) and (b) show the result of the Gaussian white{-}noise potential at $d\to\infty$ for comparison.
As $d$ increases, the Lyapunov exponent approaches a constant value. At $\omega\gg\Omega$, the average of the Lyapunov exponents in the two models coincide, in agreement with the Thouless relation $\Braket{\gamma(\omega)}=[2\ell_\pi(\omega)]^{-1}$. At $\Omega\ll\omega\ll\omega_{\rm cr}$, the difference in the Lyapunov variance between the brute-force numerics and the result for a model with $\Sigma=1$~\cite{Lifshitz1988} is attributed to insufficient medium's length. At $\omega\ll\Omega$, the average exponent and its variance saturate to different values in the two models.}}
 \end{figure}

The localization properties of a disordered 1D medium are frequently characterized by its Lyapunov exponent, $\gamma(\omega)=-(2d)^{-1}{\ln T(\omega)}$ at large $d$. In contrast with the transmission, its logarithm is indeed a self-averaging quantity~\cite{Lifshitz1988}. In Fig.~\ref{fig:lyapunov}(a) we show the frequency dependence of the averaged Lyapunov exponent, $\Braket{\gamma(\omega)}$, at fixed medium's length. Its inverse defines the localization length, $L_{\rm loc}(\omega)=1/\Braket{\gamma(\omega)}$. We checked numerically that a celebrated Thouless relation~\cite{Thouless1973}, $L_{\rm loc}(\omega)=2\ell_\pi(\omega)$, works at large frequencies, $\omega\gg \Omega$, and perfect impedance matching. 
The self-averaging nature of $\gamma$ is associated with the log-normal character of the distribution of the transmission. The frequency dependence of the ratio {$\Sigma(\omega)=d\braket{\braket{\gamma^2(\omega)}}/\braket{\gamma(\omega)}$} is plotted in Fig.~\ref{fig:lyapunov}(b). According to the Fokker-Planck formalism~\cite{Lifshitz1988}, that ratio should be 1 at $\Omega\ll \omega\ll \omega_{\rm cr}$. Despite the tendency as $d$ increases, the agreement is not perfect. (We attribute it to insufficient length.)
There was a renewed interest in $\Sigma(\omega)$~\cite{Deych2000,Schomerus2002,Deych2003,Texier2020} to test the single-parameter scaling hypothesis of the theory of localization~\cite{Abrahams1979}.

\begin{figure}
\includegraphics[width=0.9\columnwidth]{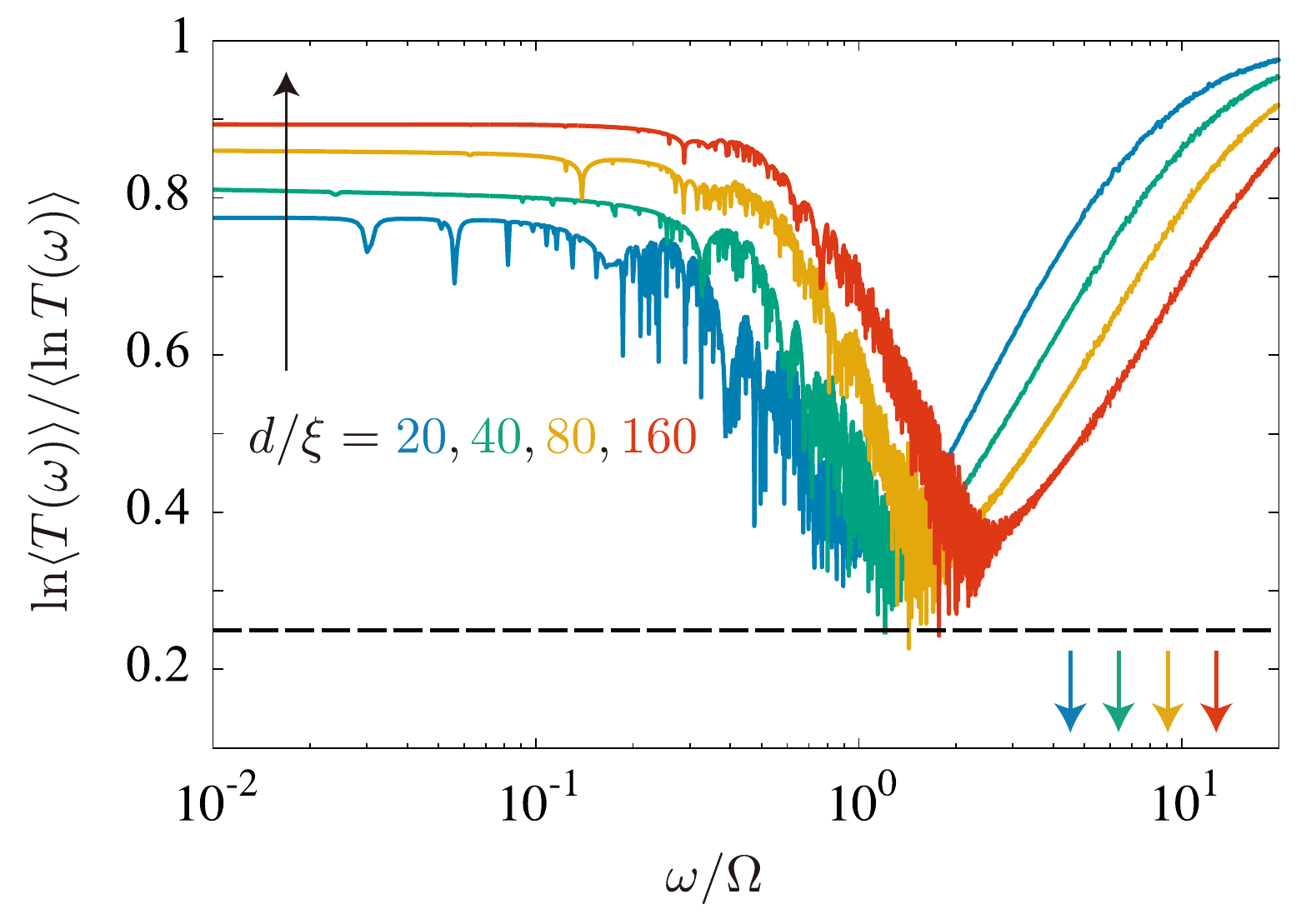}
\caption{\label{fig:ratio-lnT}
The frequency dependence of the ratio of $\ln\Braket{T(\omega)}$ and $\Braket{\ln T(\omega)}$ for different lengths $d/\xi=20,40,80,160$ and $K/K_0=1.0$. The ratio takes a frequency-independent value at low frequency, $\omega\ll\Omega$. Upon increasing the frequency, it decreases while remaining above the value $1/4$ (dashed line) predicted in the frequency range $\Omega\ll\omega\ll\omega_{\rm cr}$, then it increases to the value 1 at larger frequencies, when disorder becomes irrelevant. {Arrows indicate $\omega_{\rm cr}=\Omega\sqrt{d/\xi}$ for each of the four curves.} A scaling analysis with the length {(see footnote~\cite{scaling})} gives the result $\ln\Braket{T(\omega)}/\Braket{\ln T(\omega)}\approx 0.90$ at {$\omega\ll\Omega$ and} $d/\xi \to \infty$.} 
\end{figure}

The presence of the factor $1/4$ in the argument of the exponent in Eq.~\eqref{eq:Gertshenstein} or \eqref{eq:transmission} reflects that the averaged transmission is dominated by rare, optimal disorder configurations -- not captured by the log-normal distribution (tails), and which produce a resonant transmission $T\sim 1$~\cite{Lifshits1979,Larkin1987,Pendry1987}. In other words, $\Braket T\gg T_{\rm typ}$ where $T_{\rm typ}=\exp(\Braket {\ln T})$ is the typical transmission. To check this effect, in Fig.~\ref{fig:ratio-lnT} we plot the frequency dependence of $\ln\Braket{T(\omega)}/\Braket{\ln T(\omega)}$ for various lengths and $K=K_0$. (We attribute to insufficient length the deviation of that ratio from 1/4 at $\omega\gg\Omega$.)

{For completeness,} we compare the effect disorder has on the waves' transmission with that it has on their reflection. We may use Eq.~\eqref{eq:r} and the flat distribution of $\theta_1$ mentioned above Eq.~\eqref{eq:Tavbal} to find that the real part of the reflection amplitude averages to a frequency-independent value,
\begin{align}{
\Braket{r'(\omega)}=-r_0},
\label{eq:refl-class}
\end{align}
when $\Omega\ll\omega\ll\omega_{\rm cr}$. The averaged reflection amplitude is insensitive to disorder, in contrast to the averaged transmission coefficient $\Braket{T}$, which is manifestly dependent on $\ell_\pi$, cf.~Eqs.~\eqref{eq:Gertshenstein} and \eqref{eq:transmission}. At higher frequency, $\omega_{\rm cr}\ll \omega\ll \omega_{\rm cr}/T_0$, and $T_0\ll 1$, we use the same methods as in Sec.~\ref{sec:Ttun} to find
\begin{align}
\label{eq:rav}
\Braket{r'(\omega)}=-1+\frac{T_0}2\sum_nP_1\left(\frac{2\omega_nd}v\right).
\end{align}
It expresses the inhomogeneous broadening of plasmon standing waves confined in the medium, in correspondence with Ref.~\cite{Houzet2019,footnote}. Finally, the standard Fabry-P\'erot formula for the reflection is recovered at $\omega\gg \omega_{\rm cr}/T_0$ when the levels's broadening is dominated by the radiation to waveguides, rather than inhomogeneous broadening.

\section{Strongly localized regime (low frequency)}
\label{sec:localized}

At $\omega\ll\Omega$ plasmons are localized over a typical length $\xi$, in the fluctuations of the effective disorder potential \eqref{eq:potential}. Those fluctuations are non-Gaussian, and we have not been able to make an analytical theory of the scattering properties in that regime. However, we could gather several pieces of information from our numerics. {We compare it with known analytical results (collected in Appendix~\ref{App:gaussian}) for the 1D theory of localization in a Gaussian white{-}noise potential for which the function $w(x/\xi)$ in Eq.~\eqref{eq:correlator} is replaced by zero. While the scattering properties of the two models coincide in the weakly localized regime, $\omega\gg \Omega$, important deviations are found in the strongly localized regime, $\omega\ll\Omega$. {We show that these deviations cannot be reproduced by a model of a Gaussian colored disorder with the values of the average $\Braket{\cal V}$ and of the second cumulant $\Braket{\Braket{{\cal} V(x){\cal} V(0)}}$ read off the numerics for the pinning model, see Eq.~\eqref{eq:correlator} and Figs.~\ref{fig:potential} and  \ref{fig:potential-mean}.}

\subsection{Lyapunov exponent}
\label{sec:lyapunov}

The Lyapunov exponent remains a self-averaging quantity, which saturates to a frequency-independent value, ${\Braket{\gamma}/\xi^{-1}} \approx 1.36$ at {low} frequency. That value is larger than the one in the {Gaussian white{-}noise model}, ${\Braket{\gamma}/\xi^{-1}} =3^{1/3}\sqrt{\pi}/\Gamma(1/6)\approx 0.46 $~\cite{Derrida1984}, cf.~Fig.~\ref{fig:lyapunov}(a). 
{On the other hand, the Gaussian colored model produces a very close value for $\Braket{\gamma}$ (see Appendix~\ref{App:gaussian-colored} for details on the numerical implementation of that model), which is hardly distinguishable from the pinned-model result at any $\omega$. This raises the question whether spatial correlations dominate over non-Gaussian correlations in the determination of the plasmons' scattering properties in the strongly localized regime. Actually, our subsequent results show that both (spatial and non-Gaussian) correlations are relevant. We will argue at the end of this subsection that the similarity between the results of the pinned and Gaussian colored models for $\Braket{\gamma}$ may be fortuitous.}

In agreement with {the central limit theorem}, the variance of the Lyapunov exponent scales inversely with $d$. The frequency dependence of $\Sigma(\omega)={d}\braket{\braket{\gamma^2(\omega)}}/\braket{\gamma(\omega)}$ is plotted in Fig.~\ref{fig:lyapunov}(b). The low-frequency result, $\Sigma\approx 0.218$, is significantly smaller than the one in the {Gaussian white{-}noise model} at vanishing frequency~\cite{Schomerus2002}, $\Sigma\approx 1.1$ (that later value being quite close to the one at large frequency, $\Sigma= 1$){. The Gaussian colored model gives $\Sigma \approx 0.46$, which is also larger than the result of the pinned model. Thus a Gaussian disorder (whether it is white or colored) does note reproduce the right value of $\Sigma$, emphasizing the role of non-Gaussian correlations in the effective potential created by pinning.}

The presented values of $\Braket{\gamma}$ and $\Sigma$  were obtained by a brute-force numerical solution of the dynamical plasmon propagation problem. Next, we  show that these values are reproduced by solving the plasmon transmission coefficient problem for the static correlated random potential, Eq.~\eqref{eq:potential}, upon averaging over its realizations. According to problem \S 25.5 in Ref.~\cite{Landau-QM}, the low-frequency transmission at perfect impedance matching ($T_0=1$) for the wave equation~\eqref{eq:schroedinger} is
\begin{align}
\label{eqT0}
T(\omega\to 0)\approx \frac{4\omega^2}{v^2[\psi_0'(d)]^2}.
\end{align}
Here $\psi_0$ is the (real) solution of Eq.~\eqref{eq:schroedinger} at $\omega=0$ that satisfies the boundary conditions $\psi_0(0)=1$ and $\psi'_0(0)=0$. Equation~\eqref{eqT0} works at sufficiently small frequency for typical disorder configurations such that $\psi_0'(d)\neq 0$. [Note that $\psi_0'(d)=0$ would signal a zero-energy bound state, and a resonant transmission not captured by Eq.~\eqref{eqT0}.] An explicit solution reads
\begin{align}
\psi_0(x)=\exp\left({\xi^{-1}}\int_0^{x} dy Z(y)\right)
\end{align}
where {$Z={\xi}\psi'/\psi$ is another Ricatti variable, which solves an equation derived from Eq.~\eqref{eq:schroedinger} at $\omega=0$,}
\begin{align}
\label{eq:ric}
\xi \partial_x Z(x)=-Z^2(x)+{\cal V}(x)/\Omega^2\quad\mathrm{with}\quad Z(0)=0.
\end{align}
With this Eq.~\eqref{eqT0} reads
\begin{align}
\label{eqT0-1}
T(\omega\to 0)=4\left(\frac \omega\Omega\right)^2 \frac1{Z^{2}(d)}\exp\left(-2{\xi^{-1}}\int_0^{ d} dy Z(y)\right).
\end{align}
It was argued in Refs.~\cite{Gurarie2002,Gurarie2003} that the solution of Eq.~\eqref{eq:ric} with $\cal V$ of Eq.~\eqref{eq:potential} has a positive mean, $\Braket{Z(x)}>0$. Thus solving Eq.~\eqref{eq:ric} provides a way to find the Lyapunov exponent and its variance {from the argument of the exponential factor in Eq.~\eqref{eqT0-1}}, which only depends on $\cal V$, as announced above. This yields $\Braket{\gamma}\xi=1.37$ and $\Sigma=0.217$, in very close agreement with the results of the brute-force numerical evaluation of scattering amplitudes at a low, finite frequency{, $\Braket{\gamma}\xi=1.36$ and $\Sigma=0.218$, respectively}. The agreement is also consistent with the more general Eq.~\eqref{eq:gamma-Z} at $\omega=0$.

{Let us now return to the issue of the closeness of $\Braket{\gamma}$ evaluated within the pinned and Gaussian colored models. Using Eq.~\eqref{eq:ric}, it is straightforward to evaluate perturbatively the leading-order correction to $\Braket{\gamma}$ in the strongly localized regime, taking $\Omega/\sqrt{\Braket{\cal V}}$ as a small parameter. The result,
\begin{equation}
\label{eq:Lyap-pert}
\Braket{\gamma}\xi  \approx\frac{{\Braket{{\cal V}}}^{\frac12}}{\Omega}-\frac{\Omega^2}{2\Braket{{\cal V}}}\left[1-\frac14\int dx w(x) e^{-2|x| {\Braket{{\cal V}}}^{\frac12}/\Omega}\right]
\end{equation}
(see Appendix~\ref{App:gaussian-colored} for details on the derivation), only depends on the average and second moment of the disorder potential. Evaluating Eq.~\eqref{eq:Lyap-pert} with $\Braket{\cal V}/\Omega^2\approx 2.44$ (see Appendix~\ref{app:numerics}) yields $\Braket{\gamma}/\xi^{-1} \approx1.41$, very close to the pinned-model result, $\Braket{\gamma}/\xi^{-1} \approx1.36$. We believe that fortuitous closeness between the two results is the reason why non-Gaussian correlations seem to play a minor role in the determination of $\Braket{\gamma}$.}

\subsection{Reflection amplitude}

\begin{figure}
\includegraphics[width=0.9\columnwidth]{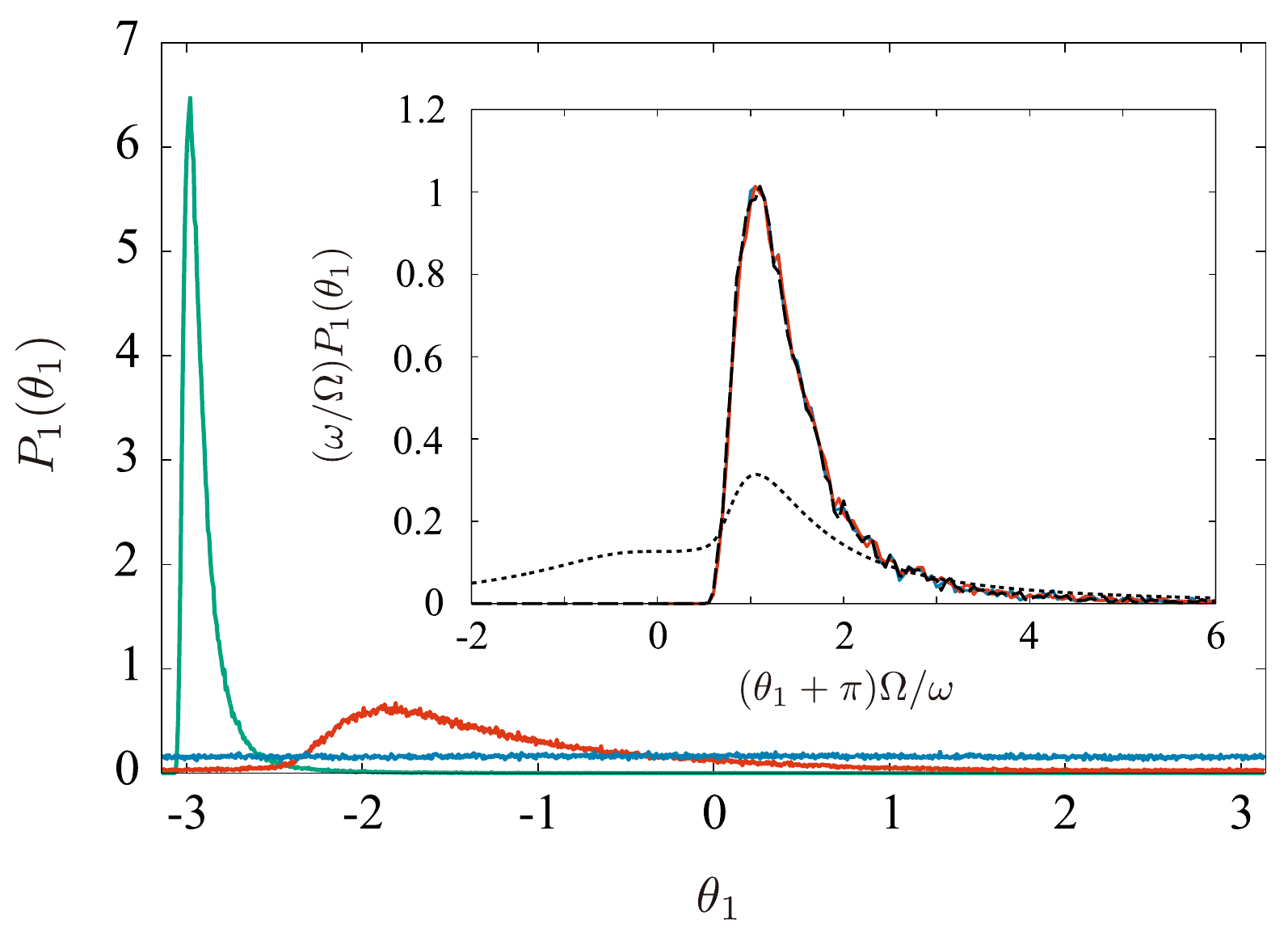}
\caption{\label{fig:proba-reflection}
Probability distributions of the reflection phase at $K/K_0=1.0$ and $d/\xi=20$, at $\omega/\Omega=$ 0.157 (green), 1.16 (red), and 8.58 (blue).
The inset {shows the universal frequency dependence of the distribution found numerically in the vicinity of phase $\theta_1=-\pi$ either by the brute force numerics at $\omega/\Omega=$ 0.1 (blue) and 0.25 (red), or from the solution of the Ricatti Eq.~\eqref{eq:ric} (dashed line).}
The dotted line represents the analytical formula derived in Ref.~\cite{Sulem1973} for a Gaussian white{-}noise disorder.} 
\end{figure}

\begin{figure}
\includegraphics[width=0.9\columnwidth]{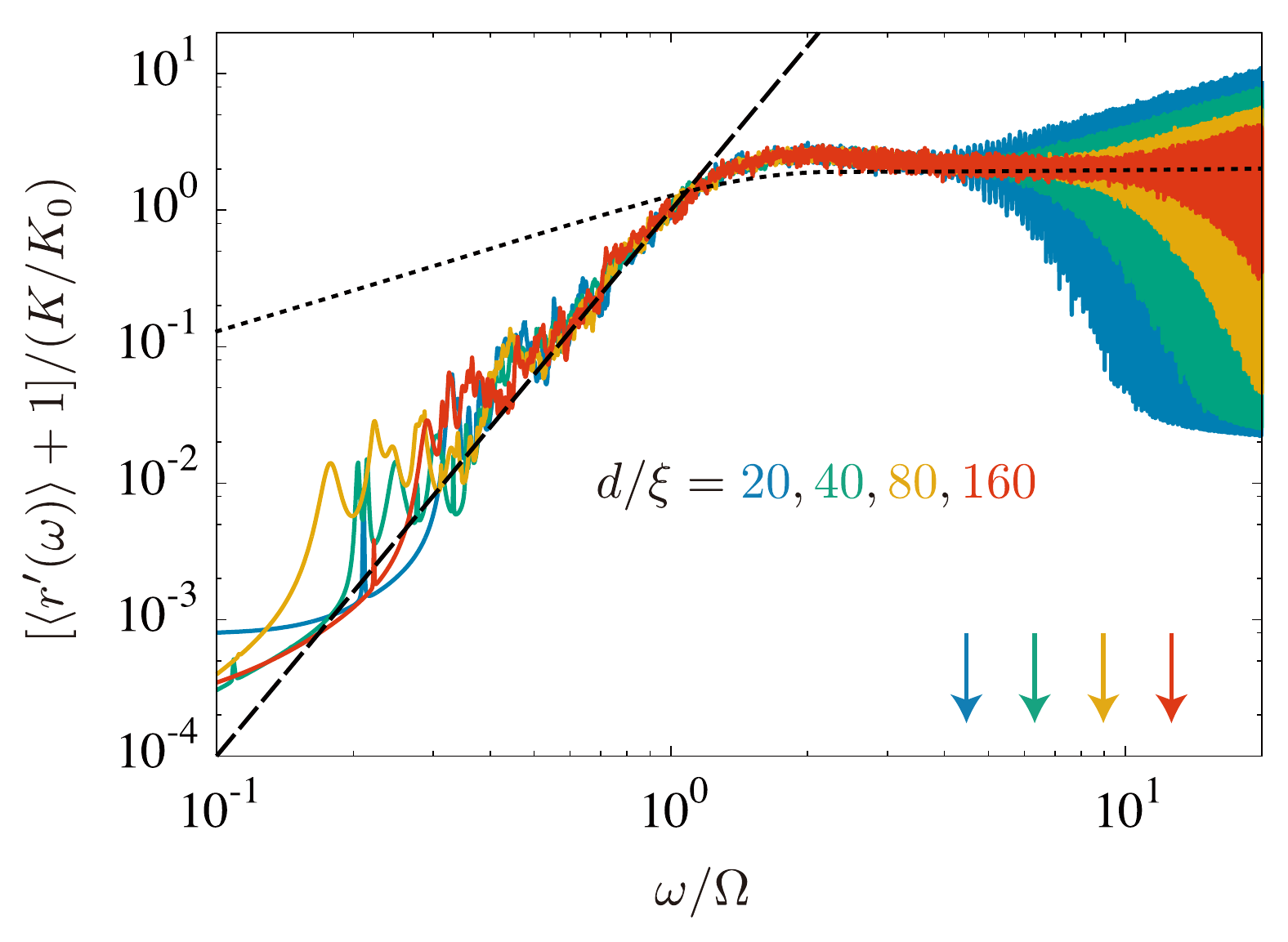}
\caption{\label{fig:DOS}
The local density of states as a function of the frequency for $K/K_0=0.01$. 
The dashed line represents $(\omega/\Omega)^4$. {[In units used in Ref.~\cite{Houzet2019}, it corresponds to $C(\omega/\omega_\star)^4$ with $\omega_\star=\Omega/(2\pi^2)^{1/3}$ and $C\approx 0.039$ (close to $C\approx 0.032$ found in Ref.~\cite{Houzet2019}).] The dotted line is the result of the Gaussian white{-}noise model, see Appendix \ref{App:gaussian}}. {Arrows indicate the crossover frequency to the ballistic regime, $\omega_{\rm cr}=\Omega\sqrt{d/\xi}$, for each $d$.} 
}
\end{figure}

{In the low-frequency regime}, the transmission is exponentially suppressed and waves are almost perfectly reflected.} In the limit of perfect impedance matching, we characterize the reflection amplitudes with the distribution of the reflection phase~\cite{Sulem1973,Barnes1990}, see Fig.~\ref{fig:proba-reflection}. While the distribution is uniform at $\omega\gg\Omega$, a single-peak structure near $\theta_1={\pm}\pi$ develops at lower frequency. 
{At $\omega\ll\Omega$, {the peak in} the distribution obtained from the brute-force numerics takes a universal frequency dependence both in our model and in the Gaussian white{-}noise model, see inset of Fig.~\ref{fig:proba-reflection}, with important differences between the two models. Furthermore, the universal dependence in our model agrees with the one obtained by solving numerically Eq.~\eqref{eq:ric} in the interval $0<x<d$ with a given (real) boundary condition at $x=d\gg\xi$ for various disorder configurations, then identifying $Z(0)={-2(\omega/\Omega)}/(\theta_1+\pi)$, which is the consequence of the boundary conditions \eqref{eq:BC1} and \eqref{eq:BC2} with $K=K_0$, $\omega/\Omega\to 0$, and $r=e^{i\theta_1}$. The results don't depend on the boundary condition for $Z(d)$ if $d\gg\xi$. The result shown in the inset of Fig.~\ref{fig:proba-reflection} is for $d/\xi=20$.}

{The distribution $P_1(\theta_1)$ can be seen as a measure of the modes localized nearby an end of the pinned region and whose frequency lies either above ($\theta_1>-\pi$) or below ($\theta_1<-\pi$) the one of the incoming wave.} 
The distribution appears to reach zero at $\theta_1=-\pi$, unlike {both the white-noise and colored Gaussian cases (see Appendices~\ref{App:gaussian-colored} and \ref{App:gaussian}, respectively)}, indicating the scarcity of low-frequency resonances. This agrees with the strong suppression of the low-frequency modes' density of states~\cite{Fogler2002,Gurarie2002,Gurarie2003} in the pinned model, in contrast with the Gaussian {models.}

{The distribution $P_1$, together with Eqs.~\eqref{eq:BC1} and \eqref{eq:BC2}, allows finding the statistics of the reflection phase at any impedance mismatch between the waveguide and half-infinite medium.} 
In particular, 
\begin{equation}
\label{eq:Imr-mismatch}
\Braket{r'}+1=\int d\theta_1P_1(\theta_1)\frac 2{1+(K_0/K)^2\tan^2(\theta_1/2)}.
\end{equation}
At $\omega\ll\Omega$, the $\omega$-scaling of the peak in $P_1$ near $\theta_1=-\pi$, demonstrated in the inset of Fig.~\ref{fig:proba-reflection}, yields{~\cite{footnote-reflection}}
\begin{equation}
\label{eq:Imr-mismatch-1}
\Braket{r'}+1=A \left(\frac K{K_0}\right)^2 \left(\frac \omega{\Omega}\right)^2, \qquad A\approx 2.4
\end{equation}
At any $K\ll K_0$, another important contribution to Eq.~\eqref{eq:Imr-mismatch}, which scales linearly with $K/K_0$, comes from angles near $\theta_1=0$, yielding
\begin{equation}
\label{eq:Imr-mismatch-large}
\Braket{r'}+1=4\pi\frac{K}{K_0}P_1(\theta_1=0).
\end{equation}
Our numerics for $\Braket{r'}$ at $K\ll K_0$, summarized by Fig.~\ref{fig:DOS}, confirms the result of Ref.~\cite{Houzet2019} and Eq.~\eqref{eq:Imr-mismatch-large} with $P_1(\theta_1=0)\propto \omega^4${; in this regime $(\Braket{r'}+1)K_0/K$ can be interpreted as the local density of plasmon modes}. However, we were not able to correlate this result with the result of a direct evaluation of $P_1(\theta_1=0)$ illustrated by Fig.~\ref{fig:proba-reflection}, presumably due to an insufficient number of disorder configurations. {Furthermore, the deviation of the numerics from the $\omega^4${-}scaling seen in Fig.~\ref{fig:DOS} at the lowest frequencies is suggestive of a crossover to the $\omega^2${-}scaling of Eq.~\eqref{eq:Imr-mismatch-1}; our numerics did not allow us to check this scaling quantitatively.} The different dependence of the r.h.s.~of Eq.~\eqref{eq:Imr-mismatch-large} on $K/K_0$, compared with Eq.~\eqref{eq:Imr-mismatch-1}, ensures that it dominates over a wide frequency range. {The scarcity of low-frequency modes reflects in a much stronger suppression of the local density of states $\propto \omega^4$ at $\omega\ll\Omega$ in the pinned model, in contrast with the results of the white-noise and colored Gaussian models where  $(\Braket{r'}+1)K_0/K\propto 1.29 \omega/\Omega$ and $0.35\omega/\Omega$, respectively (see Appendices~\ref{App:gaussian} and \ref{App:gaussian-colored}).}

{Finite-range correlations break the translational invariance of the effective disorder potential near the edges of the medium, as exemplified in Fig.~\ref{fig:potential-mean}. This prevents one from evaluating the Lyapunov exponent with Eq.~\eqref{eq:gamma-Z} together with the distribution of the Ricatti variable at the edge, which determines $P_1(\theta_1)$ plotted in Fig.~\ref{fig:proba-reflection}. Expectedly, that procedure yields $\Braket{\gamma}\xi=1.59$, different from $\Braket{\gamma}\xi\approx 1.37$ found in Sec.~\ref{sec:lyapunov}.}

\subsection{{Average transmission coefficient}}
\label{sec:transmission-av}

At any frequency in the range $\omega\ll\omega_{\rm cr}$, the averaged transmission is determined by the optimal configurations. Furthermore, the ratio $\ln \braket{T}/\braket{\ln T}$ remains smaller than 1, in accordance with the inequality between the arithmetic and geometric means. In Sec.~\ref{sec:disc}, we found $\ln \braket{T}/\braket{\ln T}=1/4$ in the diffusive regime ($\omega\gg\Omega$). In the strongly localized regime, this ratio increases towards the value $\ln \braket{T}/\braket{\ln T}\approx 0.9${, see Fig.~\ref{fig:ratio-lnT}}. Its closeness to 1 hints to a small difference between the typical and optimal disorder configurations. This is further confirmed by the fact that ensemble-averaging of Eq.~\eqref{eqT0} yields a very close result for that ratio, $\ln \braket{T}/\braket{\ln T}\approx {0.93}$. As {disorder configurations with nearly-perfect transmission~\cite{Lifshits1979}} are not taken into account in Eq.~\eqref{eqT0}, {they should have a} negligible weight among the optimal disorder configurations.

Note that the frequency scaling $\Braket{T(\omega\to 0)}\propto\omega^2$ in Eq.~\eqref{eqT0-1} confirms that, for the waves at the bottom of the plasmon spectrum (i.e., in the infinite wavelength limit), the disorder potential acts like a localized one for each realization of the disorder. Our numerics does not have enough accuracy to check this asymptote quantitatively and to establish its range of validity.

{For comparison, we expect disorder configurations with nearly-perfect transmission to play a major role at arbitrarily low frequency in the Gaussian {models}. This may result in a different frequency dependence of $\braket{T(\omega)}$ and $\ln \braket{T(\omega)}/\braket{\ln T(\omega)}$ at $\omega\ll \Omega$. As far as we know, this question has not been studied systematically.} {Our numerics indicates $\ln \braket{T(\omega)}/\braket{\ln T(\omega)}\approx 0.72$ and $\approx 0.87$ at $\omega\ll \Omega$ in the Gaussian white-noise and colored model, respectively. We attribute (again) the difference of these results from the pinned model result, $\ln \braket{T(\omega)}/\braket{\ln T(\omega)}\approx 0.9$, to the scarcity of low-frequency quasi-localized modes in the latter.} 
 
\section{Conclusion}
\label{sec:discussion}

The study of a variety of 1D models has been influential in the understanding of waves' localization in random media. A wealth of analytical results could be obtained thanks to advanced mathematical methods, most (if not all) of them relying on a Markovian assumption for the disorder, affecting the propagation of linear waves. This assumption breaks down in the case of disorder associated with the pinning of an elastic medium. Our study of the wave scattering by the pinned elastic medium reveals the universality of the scattering properties, despite the manifest breakdown of the Markovian nature of disorder. 

The spatial correlations of the pinning-induced disorder affect the wave propagation at all frequencies. In the high-frequency regime, we could use the familiar {stochastic} methods for Gaussian white{-}noise disorder, {see Sec.~\ref{sec:FP}. Surprisingly, we found that the correlations still affect the forward-scattering length, despite the wavelength being shorter than the Larkin length in that regime, see Eqs.~\eqref{eq:lengths} and \eqref{eq:mfp}. In the almost-ballistic regime (mean-free path exceeding the medium's length) and at a strong impedance mismatch between the pinned medium and the waveguides, the transmission coefficient $\Braket{T}$ exhibits resonances, inhomogeneously broadened by the disorder. Correlations affect the broadening, see Eqs.~\eqref{eq:Ttun-cross}, \eqref{eq:FP_P1-sol}, and Fig.~\ref{fig:T_semiclass}(b).}

On the other hand, at lower frequency, the wave localization is strong, {the stochastic} methods relying on the Fokker-Planck equation obviously do not work. In the strong localization regime, our findings, which are discussed in Sec.~\ref{sec:localized}, are mostly numerical. We found that the Lyapunov exponent is significantly increased, and its variance is suppressed, in comparison with a Gaussian white{-}noise disorder of a comparable strength, see Fig.~\ref{fig:lyapunov}. 
{A large part of the increase of the Lyapunov exponent can be accounted by the spatial correlations on the Larkin length in the pinned model of disorder; on the other hand, both the spatial and non-Gaussian correlations contribute to the Lyapunov's variance suppression.} 
The closeness of the ratio $\ln \braket{T(\omega)}/\braket{\ln T(\omega)}$ to 1, see Fig.~\ref{fig:ratio-lnT}, allows us to infer a similarity between the optimal for the transmission and the typical disorder configurations. This is in strong contrast with the diffusive regime, in which the same ratio is significantly suppressed, $\ln \braket{T(\omega)}/\braket{\ln T(\omega)}=1/4$, as we discussed in Sec.~\ref{sec:disc}. In the regime of strong localization, in which the transmission coefficient is exponentially suppressed with the medium's length, we also found signatures of the scarcity of the localized low-frequency plasmon modes. This mode scarcity affects the distribution function of the phase of reflection off the pinned medium, see Figs.~\ref{fig:proba-reflection} and \ref{fig:DOS}. {These observables yield unique signatures of the non-Gaussianity of the pinned medium, such as the suppression of the reflection phase distribution  $P_1(\theta_1=-\pi)$  and an average transmission probability through the medium that is much closer to the typical one, than for the Gaussian models of disorder.} {Connections between the static pinning theory and the localization theory have been recently explored in Ref.~\cite{Fyodorov2018}.} We hope that our study provides a momentum for further developments of the theory aiming at better understanding of wave dynamics in a pinned elastic medium.

Furthermore, our focus was mostly on the collective pinning regime, i.e., Larkin length long compared to the spacing between the sites comprising the medium. The case of strong pinning may also reveal new physics.

Finally, we argued that Josephson-junction arrays provide a versatile system to address the interplay of elasticity and disorder by their microwave spectroscopy. Here we would like to stress that the development of superinductances~\cite{Manucharyan2009} allowed reaching a regime where $K$ is small, but not vanishingly small. Therefore it would be interesting to investigate the deviations from classical theory, especially the inelastic scattering phenomena that will be induced by quantum fluctuations~\cite{Bard2017,Houzet2019}.

\begin{acknowledgments}
We acknowledge interesting correspondence with P.~Le Doussal, C.~Texier{, and M. Filippone}.
The authors thank the Supercomputer Center, the Institute for Solid State Physics, the University of Tokyo for the use of the facilities. The work was supported by Grant-in-Aid for JSPS Fellows Grant Number 20J11318 (TY), by DOE contract DEFG02-08ER46482 (LG), and by ANR through Grant No.~ANR-16-CE30-0019 (MH).
\end{acknowledgments}

\appendix

\section{Numerics}
\label{app:numerics}

Here we describe our strategy to calculate the scattering amplitudes numerically.

\subsection{Effective disorder potential}

First, we rewrite our problem with the dimensionless variables by rescaling the spatial dimension by $\xi$ and the time dimension by $1/\Omega$.
Next, the dimensionless spatial variables are discretized with a small spacing $\epsilon$, which divides the medium into $M$ intervals, $d/\xi=(M+1)\epsilon$.
Then, the above prescription leads the classical energy functional~\eqref{eq:E}, in units of $({\epsilon \hbar} v/2\pi K\xi)$, to take the discretized form:
\begin{align}
   \label{eq:E_dis}
   &\tilde{\cal E}[\{\theta_m\}]
   =\sum_{m=0}^{M-1}\left[\left(\frac{\theta_{m+1}-\theta_m}{\epsilon}\right)^2
   -\frac{{\cal V}_m}{2}\right]-\frac{{\cal V}_0+{\cal V}_M}{4}.
\end{align}
Here, 
\begin{align}
\label{eq:effective}
{\cal V}_m={2}\left[{\cal V}'_m\cos2\theta_m+{\cal V}''_m\sin2\theta_m\right]
\end{align} 
is the effective disorder potential in dimensionless units, and the correlators \eqref{eq:disorder} are reproduced in continuum limit, $\epsilon\to 0$, if $\Braket{{\cal V}'_m}=\Braket{{\cal V}''_m}=\Braket{{\cal V}'_m{\cal V}''_n}=0$ and $\Braket{{\cal V}'_m{\cal V}'_n}=\Braket{{\cal V}''_m{\cal V}''_n}=(1/\epsilon)\delta_{m,n}$.
The last term in Eq.~\eqref{eq:E_dis} arose from considering an infinite periodic array with the mirror symmetry ${\cal V}_m={\cal V}_{2M-m}$ in unit array ($0\le m\le 2M-1$). 
This ensures that the boundary conditions $\partial_x\theta(0)=\partial_x\theta(d)=0$ are satisfied in continuum limit. 

We compared two ways to generate the random fields. In the ``phase model'' reminiscent of the original model, we relate them with flatly distributed random phases at each site, ${\cal V}'_m=\sqrt{ 2/\epsilon}\cos \chi_m$ and ${\cal V}''_m=\sqrt{ 2/\epsilon}\sin \chi_m$ with $0<\chi _m<2\pi$. Alternatively, in the ``box model'', ${\cal V}_m'$ and ${\cal V}_m''$ are independent and flatly distributed in the interval $-\sqrt{3/\epsilon}<{\cal V}'_m,{\cal V}_m''<\sqrt{3/\epsilon}$. Our numerics could not distinguish between the two models.
 
The static charge density is determined by minimizing the classical energy functional~\eqref{eq:E_dis}.
For this we adopted the minimization method described in Ref.~\cite{Gurarie2003}. Note that a common $\pi$-shift of the angles, $\theta_m\to\theta_m+\pi$, leaves Eq.~\eqref{eq:E_dis} invariant; furthermore, the effective potential \eqref{eq:effective} only depends on $\bar\theta_m\mod \pi$. Thus we can look for solutions such that $-\pi/2<\bar\theta_m\leq \pi/2$, provided we make the substitution 
$(\theta_m-\theta_{m+1})^2\to \min_k(\theta_m-\theta_{m+1}-k\pi)^2$ in Eq.~\eqref{eq:E_dis}.
({At weak pinning, it is enough to keep $k=-1,0,1$.})

\begin{figure}
\includegraphics[width=0.9\columnwidth]{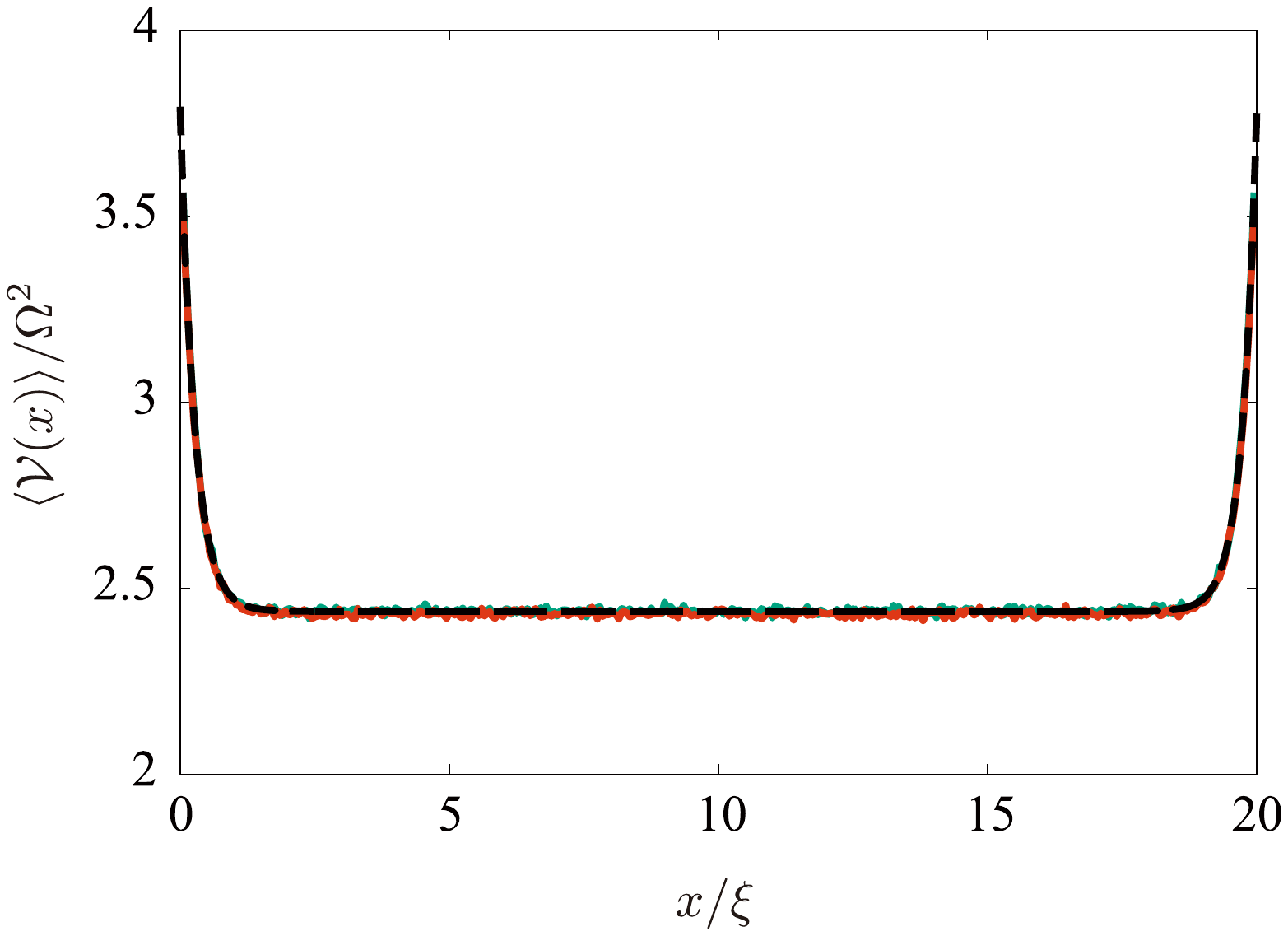}
\caption{\label{fig:potential-mean}
{The mean effective disorder potential} at $d/\xi=20$.
The green and red lines are the numerically obtained curves for {$10^4$} realizations of the random ``phase'' and ``box'' models (see Appendix~\ref{app:numerics}), respectively; they cannot be distinguished from each other. The dashed lines represent the fitting functions $v(x)=d[e^{-x/b}+e^{-(d-x)/b}]+v_0$ ($d=1.36,~b=0.269$, $v_0=2.44$).} 
\end{figure}

Generating $10^4$ random configurations at fixed medium's length to calculate the static charge density for each of them, we can evaluate the mean and second moment of the disorder potential \eqref{eq:potential}.
The results are shown in Figs.~\ref{fig:potential-mean} and \ref{fig:potential}, respectively.
The mean reaches a constant value sufficiently far from the edges, on the scale of $\xi$.
We also found numerically a fitting function that describes the enhancement of $\Braket{{\cal V}(x)}$ near the edges, see the legend of Fig.~\ref{fig:potential-mean}. It works well at any $d>20\xi$ (not shown).
We also found numerically that the long-range correlation in the second moment of ${\cal V}(x)$ are well described by Eq.~\eqref{eq:correlator} with the exponentially decaying function $w(x)=ce^{-|x|/a}$. Here $c\approx3.24$ and $a\approx0.346$, cf.~Fig.~\ref{fig:potential}. 

\subsection{Scattering amplitudes}

The wave equation~\eqref{eq:schroedinger} and boundary conditions~\eqref{eq:BC} are discretized as
\begin{align}
    \label{eq:schroedinger_dis}
    \left(\frac{\omega}{\Omega}\right)^2\psi_m+\frac{\psi_{m+1}+\psi_{m-1}-2\psi_m}{\epsilon^2}-{\cal V}_m\psi_m=0,
\end{align}
for $0<m\leq M$, and 
\begin{subequations}
\label{eq:BC-discrete}
\begin{align}
    1+r(\omega)&=\psi_0,\\
    i\omega{K}/{K_0}(1-r(\omega))&=\Omega(\psi_1-\psi_0)/\epsilon, \\
    t(\omega)&=\psi_M, \\
    i\omega{K}/{K_0}t(\omega)&=\Omega(\psi_M-\psi_{M-1})/\epsilon,
\end{align}
\end{subequations}
respectively.
The boundary conditions \eqref{eq:BC-discrete}
are used to express $r(\omega),t(\omega),\psi_0,\psi_M$ in terms of $\psi_1,\psi_{M-1}$. Inserting the found $\psi_0,\psi_M$ into Eq.~\eqref{eq:schroedinger_dis}, we find that the latter forms a linear eigenvalue with a source term on the vector $\{\psi_1,\dots,\psi_{M-1}\}$. For a given disorder configuration yielding the potentials ${\cal V}_m$, the eigenvalue problem is solved and we obtain the scattering amplitudes from relations
\begin{subequations}
\begin{align}
    r(\omega)
    &= \left(1-i\epsilon\frac{K}{K_0}\frac{\omega}{\Omega}\right)^{-1}\left(\psi_1-1-i\epsilon\frac{K}{K_0}\frac{\omega}{\Omega}\right), \\
    t(\omega)
    &= \left(1-i\epsilon\frac{K}{K_0}\frac{\omega}{\Omega}\right)^{-1}\psi_{M-1}.
\end{align}
\end{subequations}
The averaged scattering amplitudes are evaluated numerically by iterating this procedure for a large number of disorder configurations. We used $10^4$ disorder configurations to generate the plots in Figs. \ref{fig:potential}, \ref{fig:T_semiclass}, \ref{fig:proba-reflection}, and \ref{fig:potential-mean}, and $10^3$ in the plots in Figs~\ref{fig:lyapunov}, \ref{fig:ratio-lnT}, and  \ref{fig:DOS}, as well as {to compute ensemble-averages discussed in Sec.~\ref{sec:localized}, which are based on the discretized version of Eqs.~\eqref{eq:ric} and \eqref{eqT0-1}.}

\section{Fokker-Planck equation}
\label{App:FP}

In this Appendix, we provide the recipe that can be used to obtain the Fokker-Planck Eq.~\eqref{eq:FP} from the Ricatti Eq.~\eqref{eq:ricatti} and Gaussian correlators \eqref{eq:correlator-Dirac}.

We consider the system of dynamical equations
\begin{equation}
\partial_tX_i=a_i(\bm{X})+\sum_jb_{il}(\bm{X}) \xi_l(t)
\end{equation}
on the multi-dimensional variable $\bm{X}=(X_1,\dots,X_N)$ with random forces characterized by  Gaussian correlators,
\begin{equation}
\langle \xi_l(t)\xi_m(t')\rangle=2D_{lm}\delta(t-t').
\end{equation}
The joint probability $P(t,\bm{X})$ of the variables $X_1,\dots,X_N$ satisfies the Fokker-Planck equation
\begin{equation}
\frac{\partial P}{\partial t}=-\sum_i\frac{\partial}{\partial X_i}(a_iP)+\sum_{ijlm}D_{lm}\frac{\partial}{\partial X_i}\left[b_{il}\frac{\partial}{\partial X_j}(b_{jm}P)\right].
\end{equation}

\section{Exact solution of $P_2(x,\theta_2)$}
\label{App:P2}

In this Appendix, we {closely follow Ref.~\cite{beenakker1993} to} find the solution of Eq.~\eqref{eq:FP_P2} for the conditional probability that satisfies the initial condition $P_2(d,\theta_2)=\delta(\theta_2+\ln r_0)$.

For this we use various changes of variable to recover equivalent formulations of that equation. With $e^{-2\theta_2}=\lambda/(1+\lambda)$ ($\lambda>0$) and $\tilde P(\tau,\lambda)=P_2(\tau,\theta_2)|d\theta_2/d\lambda|$ (to ensure that normalization is preserved), we find
\begin{equation}
\frac{\partial \tilde P}{\partial \tau}=\frac\partial{\partial \lambda}\lambda(1+\lambda)\frac\partial{\partial \lambda}\tilde P
\end{equation}
with $\tau=(d-x)/\ell_\pi$. Introducing $\lambda=\sinh^2X$ ($X>0$) and $\bar  P(\tau,X)=\tilde P(\tau,\lambda)|d\lambda/dX|$ yields
\begin{equation}
\label{eq:FP5}
\frac{\partial }{\partial \tau} {\bar P}=\frac 14\left[\frac{\partial^2}{\partial X^2} {\bar P} -\frac{\partial}{\partial X}\left(\frac{2\cosh2X}{\sinh 2X}  {\bar P} \right)\right].
\end{equation}
Then, introducing $Q(X)= {\bar P}(X)/\sqrt{\sinh 2X}$, we find
\begin{equation}
-\frac{\partial }{\partial \tau}Q=\left( {\cal H}+\frac 14\right)Q
\end{equation}
with an effective Hamiltonian
\begin{equation}
\label{eq:Heff}
{\cal H}=-\frac 14\frac{\partial^2 }{\partial X^2}-\frac 1{4\sinh^2 2X}.
\end{equation}
Together with Eq.~\eqref{eq:FP5}, the probability conservation,
\begin{equation}
\int_0^\infty dX{\bar P}(\tau,X)=1,
\end{equation}
imposes
\begin{equation}
\lim_{X\to 0}\left(\frac{\partial \bar { P}}{\partial X}-\frac { {\bar P}}X\right)=0
\end{equation}
and, subsequently,
\begin{equation}
\label{eq:boundary}
\lim_{X\to 0}\left(\frac{\partial Q}{\partial X}-\frac {Q}{2X}\right)=0.
\end{equation}
Knowing a normalized eigenspectrum of Eq.~\eqref{eq:Heff}, ${\cal H}\psi_k=\varepsilon_k\psi_k$, together with the boundary condition \eqref{eq:boundary} then allows expressing the conditional probability
\begin{equation}
{\bar P}(\tau,X)=\sqrt{\frac {\sinh 2X}{\sinh 2X'}}\sum_k\psi_k(X)\psi^*_k(X')e^{-(\varepsilon_k+\frac14)\tau},
\label{eq:tildeP}
\end{equation}
such that $P(0,X)=\delta(X-X')$ with $\cosh 2X'=(2-T_0)/T_0$.
{(Reference~\cite{beenakker1993}} considered the case $T_0=1$, corresponding to $X'=0$, only.) The next paragraph is devoted to the solution of the eigenproblem.

The change of variable $\psi_k(X)=\sqrt{\sinh 2X}f_k(\cosh 2X)$ shows that $f_k(z)$ with $z=\cosh 2X>1$ solves the Legendre differential equation
\begin{equation}
\label{eq:legendre}
\frac{d}{dz}\left[(1-z^2)\frac{df_k}{dz}\right]+\nu(\nu+1)f_k=0
\end{equation}
with $(\nu+1/2)^2=-\varepsilon_k$. The boundary condition \eqref{eq:boundary} translates into
\begin{equation} 
\label{eq:boundary2}
\left[(z-1)^{3/4}\frac{d f_k(z)}{d z}\right]_{z\to 1}=0.
\end{equation}
Equation \eqref{eq:legendre} is solved by the Legendre functions of the first and second kind, $P_\nu(z)$ and $Q_\nu(z)$, respectively. The boundary condition \eqref{eq:boundary2} excludes $Q_\nu(z)$, which diverges logarithmically at $z\to 1$, from the set of solutions. As $P_\nu(z)\sim z^{-\nu-1}$, the normalizability condition of the wavefunctions also requires ${\rm Re}\,\nu>1/2$. Therefore, only solutions with $\varepsilon_k>0$ (meaning that the Hamiltonian \eqref{eq:Heff} defined at $X>0$ does not admit for bound states) are allowed. The set of solutions is thus given by the conical functions, $f_k(z)=C_kP_{-1/2+ik/2}(z)$ with $\varepsilon_k=k^2/4$ and $k>0$ [as $P_{-1/2+ik/2}(z)=P_{-1/2-ik/2}(z)$]. The normalization condition
\begin{equation}
\int _0^\infty dx\psi_k(x)\psi_{k'}(x)=\delta(k-k')
\end{equation}
is then obtained by choosing $C_k=\sqrt{\pi k\tanh(\pi k/2)}$.
Equation \eqref{eq:tildeP} then reads
\begin{align}
{\bar P}(\tau,X)=\frac12e^{-\frac{\tau}4}\sinh2X\int_0^\infty \!\!\!\!\!\!dkk\tanh\frac{\pi k}2
P_{-\frac12+i\frac k2}(\cosh2X)
\nonumber
\\
\times P_{-\frac12+i\frac k2}(\cosh 2X')e^{-\frac{k^2\tau}4}.
\label{eq:tildeP2}
\end{align}
Reverting to variable $\theta_2$ and setting $x=0$, i.e., $\tau=d/\ell_\pi$, we obtain Eq.~\eqref{eq:P2}.

\section{Gaussian white-noise disorder}
\label{App:gaussian}

In this Appendix we quote analytical results from the literature on a model with Gaussian white-noise disorder, which were used in plotting some of the lines in Figs.~\ref{fig:lyapunov}, \ref{fig:DOS} and \ref{fig:proba-reflection}. We assume that the variance of the disorder potential corresponds to the first term of Eq.~\eqref{eq:correlator}. Note that the literature mostly adresses the stationary Schr\"odinger equation at energy $E$. In the context of Eq.~\eqref{eq:schroedinger}, we set $E=\omega^2$ in the results from literature, constraining them to $E\geq 0$.

The Lyapunov exponent is given by~\cite{Derrida1984}
\begin{equation}
\label{eq:derrida}
\langle \gamma(\omega)\rangle\xi=\frac 12\frac{\int_0^\infty dx \sqrt{x}\exp\left(-{x^3}/{24}-{x\omega^2}/{2\Omega^2}\right)}
{\int_0^\infty dx /\sqrt{x}\exp\left(-{x^3}/{24}-{x\omega^2}/{2\Omega^2}\right)};
\end{equation}  
it is plotted by dotted line in Fig.~\ref{fig:lyapunov}(a). It yields the zero-frequency result
\begin{equation}
\langle\gamma(0)\rangle \xi =\frac{3^{1/3}\sqrt{\pi}}{\Gamma(1/6)}\approx 0.46,
\end{equation}
as well as $\langle\gamma(0)\rangle \xi=\Omega^2/2\omega^2$ at $\omega\gg\Omega$.

The Lyapunov's variance is given by {\cite{Texier2020}}
\begin{equation}
\langle\langle\gamma^2(\omega)\rangle\rangle \xi d=\int_0^\infty\frac {ds}s{\rm Re}\left[\left(2\langle\gamma(\omega)\rangle \xi-i\frac{d}{ds}\right)f^2(s)\right]
\end{equation}
with $f(s)=\phi(s)/\phi(0)$ and 
\begin{equation}
\phi(s)={\rm Ai}\left(-\frac{(\omega/\Omega)^2+2is}{2^{2/3}}\right)-i \,{\rm Bi}\left(-\frac{(\omega/\Omega)^2+2is}{2^{2/3}}\right),
\end{equation}
where Ai and Bi are Airy functions; it is plotted by dotted line in Fig.~\ref{fig:lyapunov}(b).
It yields the zero-frequency result~{\cite{Schomerus2002}}
\begin{eqnarray}
\langle\langle\gamma^2(0)\rangle\rangle d &=&
\langle\gamma(0) \rangle \left[\frac{5\pi}{3\sqrt{3}}  - \, _3F_2\left(1,1,\frac{7}{6};\frac{3}{2},\frac{3}{2};\frac{3}{4}\right)\right]\nonumber\\
&\approx& 1.1\langle\gamma(0) \rangle
\end{eqnarray}
(here $_3F_2$ is an hypergeometric function), as well as $\langle\langle\gamma^2(\omega)\rangle\rangle d=\langle\gamma(\omega)\rangle $ at $\omega\gg\Omega$.

The distribution of the reflection phase {at perfect impedance matching between a half-infinite medium and a waveguide} \cite{Sulem1973},
\begin{equation}
\label{eq:P1-gauss}
P_1(\theta_1)=\left.\frac{(\omega/\Omega)^2+Z^2}{2{|}\omega{|}/\Omega}P(Z)\right|_{Z=-(\omega/\Omega)\tan(\theta_1/2)},
\end{equation}
{is related with the distribution of the Ricatti variable $Z$,} 
\begin{equation}
\label{eq:sulem}
P(Z)=\frac{\int_0^\infty dx \exp\left(-{x^3}/{24}-{x\omega^2}/{2\Omega^2}-x/2(Z-x/2)^2\right)}
{\int_0^\infty dx\sqrt{2\pi/x}\exp\left(-{x^3}/{24}-{x\omega^2}/{2\Omega^2}\right)}.
\end{equation}
With the help of Eqs.~\eqref{eq:derrida} and \eqref{eq:sulem}, one may check that the relation
\begin{equation}
\label{eq:gamma-Z}
\langle\gamma(\omega)\rangle \xi=\int dZ Z P(Z)
\end{equation}
holds at any $\omega$.
At $\omega\gg\Omega$, the distribution is uniform, $P_1(\theta_1)\approx 1/2\pi$. At $0<\omega\ll\Omega$, the distribution is concentrated around phase $\theta_1=-\pi$, with 
\begin{eqnarray}
\label{eq:P1-approx}
P_1(\theta_1)&\approx&\frac\Omega\omega p\left(\frac{\Omega(\theta_1+\pi)}\omega\right), \\
p(\varphi)&=&\frac{3^{5/6}}{\sqrt{\pi}\Gamma(1/6)}\int_0^\infty \frac{dx}{\varphi^2}\exp\left(-x^3/6+x^2/\varphi-2x/\varphi^2\right)\nonumber
\end{eqnarray}
{if $|\theta_1+\pi|\ll 1$.} The function $p(\varphi)$ is plotted by dotted line in the inset of Fig.~\ref{fig:proba-reflection}. 
{The tails of the distribution $P_1$ at $|\theta_1+\pi |\gg \omega/\Omega$ are given by
\begin{equation}
\label{eq:P1-tail}
P_1(\theta_1)\approx\frac{\alpha\omega/\Omega}{\cos^2(\theta_1/2)},\quad \alpha=\frac{3^{1/6}\Gamma(2/3)}{2^{7/3}\pi}.
\end{equation}
Note that Eqs.~\eqref{eq:P1-approx} and \eqref{eq:P1-tail} match each other in their common range of validity, $\omega/\Omega\ll |\theta_1+\pi |\ll 1$.}

Equation~\eqref{eq:P1-gauss} allows finding the disorder average of the reflection phase's real part at any impedance mismatch $K<K_0$,
\begin{equation}
\label{Imr-mismatch}
r'+1=\frac 2{1+(K_0/K)^2\tan^2(\theta_1/2)}.
\end{equation}
At $\omega\gg\Omega$, Eq.~\eqref{eq:refl-class} is reproduced with the uniform distribution. At $0<\omega\ll\Omega$, the average is contributed by the tails of the distribution $P_1$, Eq.~\eqref{eq:P1-tail}, which yields
\begin{equation}
\label{eq:Rer-small-freq}
\Braket{r'}+1=4\pi\alpha\frac{K}{K_0}\frac\omega\Omega.
\end{equation}
At large impedance mismatch, $K\ll K_0$, Eq.~\eqref{Imr-mismatch} is approximated by $r'+1\approx 4\pi (K/K_0)\delta(\theta_1)$, such that 
\begin{equation}
\label{eq:for-fig8}
\Braket{r'}+1=4\pi\frac{K}{K_0}P_1(\theta_1=0)
\end{equation} 
with $P_1$ of Eq.~\eqref{eq:P1-gauss}, which is plotted by dotted line in Fig.~\ref{fig:DOS}.

{The linear frequency dependence in Eq.~\eqref{eq:Rer-small-freq} at low frequency, as well as the saturation at $\omega\gg\Omega$, mirror the frequency dependence of the bulk plasmon density of states. Indeed by using the particle density of states of the Schr\"odinger problem~{\cite{Frisch1960,Halperin1965,Derrida1984}} at energy $E=\omega^2$ we find} 
\begin{equation}
\label{eq:DOS-gauss}
\frac{\nu(\omega)}{\nu_0}=\frac{2\omega}\Omega\sqrt{\frac \pi 2}
\frac{\int_0^\infty dx \sqrt{x}\exp\left(-{x^3}/{24}-{x\omega^2}/{2\Omega^2}\right)}
{\left[\int_0^\infty dx/\sqrt{x}\exp\left(-{x^3}/{24}-{x\omega^2}/{2\Omega^2}\right)\right]^2},
\end{equation}
where $\nu_0=1/\pi v$. 
In particular, $\nu(\omega)=\nu_0$ at $\omega\gg\Omega$ and 
\begin{equation}
\label{eq:DOS-gauss}
\frac{\nu(\omega)}{\nu_0}=\frac{6\pi \times 3^{1/{6}}}{\Gamma^2(1/6)}\frac\omega\Omega,\qquad {0<}\omega\ll\Omega.
\end{equation}
{The linear $\omega$-dependence in Eq.~\eqref{eq:DOS-gauss} reflects the finite (particle) density of states $\tilde \nu(0)$ of the Schr\"odinger problem at $E=0$, together with the relation $\nu(\omega)=2\omega \tilde \nu(\omega^2)$, which connects the waves' and particles' densities of states. The finiteness of $\tilde \nu(0)$ (instead of its divergence in the disorder-free problem) is a precursor of the Lifshits tail of states at $E<0$ (a region inaccessible for waves).
Note that the overall frequency dependences of $\Braket{r'(\omega)}+1$ and $\nu(\omega)$ are different.}

{Note that Eqs.~\eqref{eq:derrida} and \eqref{eq:DOS-gauss} alternatively read}
\begin{subequations}
\begin{eqnarray}
\langle \gamma(\omega)\rangle\xi&=&-{\rm Im}\left[f'(s=0)\right],
\\
{\nu(\omega)}/\nu_0&=&-\Omega {\rm Re}\left[\frac d{d\omega} f'(s=0)\right],
\end{eqnarray}
\end{subequations}
respectively. {Actually, t}he relation of $\langle\gamma(\omega)\rangle$ and $\nu(\omega)$ to a common analytic function in the upper complex plane is {valid beyond the Gaussian white{-}noise model. It is}
a consequence of the Herbert-Jones-Thouless relation~\cite{Thouless1972}  for any translationally invariant potential,
\begin{equation}
\label{eq:HJTh}
\gamma(E)-\gamma_0(E)=\int dE'\ln|E-E'|\left[\tilde \nu(E')-\tilde \nu_0(E')\right].
\end{equation}
Here $\gamma_0(E)=\Theta(-E)\sqrt{-E}/v$ and $\tilde \nu_0(E)= \Theta(E)/(2\pi v \sqrt{E})$ are for a free particle. In the case of a random potential, $\gamma(E)$ and $\tilde\nu(E)$ are the respective functions averaged over the disorder realizations, and the translational invariance property refers to the correlation function of the potential. Equation~\eqref{eq:gamma-Z}, which relates the Lyapunov exponent with the average the Ricatti variable, {also holds for a translationally invariant disorder potential,} under an additional assumption of inversion symmetric disorder~\cite{Lifshitz1988}.

{\section{Gaussian colored-noise disorder}
\label{App:gaussian-colored}
}

In Sec.~\ref{Gauss-an}, we derive asymptotic formulas for the Lyapunov exponent at small and large frequency; they only depend on the average and second moment of the disorder potential. In Sec.~\ref{Gauss-num} we show how to implement a random Gaussian colored potential numerically; we provide plots that illustrate the difference between the pinned and Gaussian colored models for the plasmon's scattering properties.

\subsection{Asymptotes for the Lyapunov exponent}
\label{Gauss-an}

Let us start with the small frequency regime. The Lyapunov exponent is a self-averaging quantity defined as $\langle\gamma\rangle=Z(x\to \infty)/\xi$, where $Z(x)$ is a solution of the Ricatti equation
\begin{align}
\xi \partial_x Z(x)=-Z^2(x)+\left[{\cal V}(x)-\omega^2\right]/\Omega^2,
\end{align}
which generalizes Eq.~\eqref{eq:ric} at finite frequency, irrespective of the initial condition $Z(x=0)$. We ignore mathematical details related with the existence of the Lyapunov exponent~\cite{Lifshitz1988} and note that $Z(x\to \infty)=\Braket{Z(x\to \infty)}$.

Let us assume $\omega<\sqrt{\Braket{\cal V}}$. Then, treating $v(x)={\cal V}(x)-\Braket{\cal V}$ as a perturbation, we find that the leading-order contribution to $Z(x)=Z^{(0)}(x)+Z^{(1)}(x)+Z^{(2)}(x)+\dots$ is constant, $Z^{(0)}=\sqrt{\Braket{\cal V}-\omega^2}/\Omega$. The next-order term is found as the solution of the equation
\begin{align}
\xi \partial_x Z^{(1)}(x)=-2Z^{(0)}Z^{(1)}(x)+v(x)/\Omega^2;
\end{align}
it yields
\begin{align}
Z^{(1)}(x)=Z^{(1)}(0)e^{-2xZ^{(0)}/\xi}+\int_0^x\frac{dy}\xi\frac{v(y)}{\Omega^2}e^{-2(x-y)Z^{(0)}/\xi}.
\end{align}
As $\Braket{Z^{(1)}(x\to \infty)}=0$, it does not contribute to the Lyapunov exponent. In the next order, 
\begin{align}
\xi \partial_x Z^{(2)}(x)=-2Z^{(0)}Z^{(2)}(x)-Z^{(1)}(x)^2;
\end{align}
the solution yields
\begin{align}
\Braket{Z^{(2)}(x\to \infty)}=-\frac1{8Z^{(0)2}}\int \frac{dx}{\xi} \frac{\Braket{v(x)v(0)}}{\Omega^4}e^{-2|x|Z^{(0)}/\xi},
\end{align}
which only depends on the second cumulant of the disorder potential. Inserting Eq.~\eqref{eq:correlator} for the correlator $\Braket{v(x)v(0)}=\langle\langle{\cal V}(x){\cal V}(0)\rangle\rangle$, we find 
Eq.~\eqref{eq:Lyap-pert} at vanishing frequency.

Switching to the regime of large frequencies, we introduce $\psi=\rho \sin\vartheta$ and $\psi'=\kappa\rho \cos\vartheta$ with $\kappa=\sqrt{\omega^2-{\cal V}}/v$ to find that the Schr\"odinger equation, Eq.~\eqref{eq:schroedinger}, transforms into
\begin{equation}
\label{eq:schord-evelope}
\vartheta'=\kappa-\frac{v(x)}\kappa\sin^2\vartheta, \quad \rho'=\rho \frac{v(x)}{2\kappa}\sin 2\vartheta.
\end{equation}
The Lyapunov exponent can also be defined as the rate of growth of the wavefunction's envelope~\cite{Lifshitz1988}, 
\begin{equation}
\label{eq:lyap-envelope}
\langle \gamma\rangle=\lim_{x\to \infty}\frac{\ln \rho(x)}x.
\end{equation}
Solving Eq.~\eqref{eq:schord-evelope} perturbatively in $v(x)$ and inserting the solution into Eq.~\eqref{eq:lyap-envelope}, we find that the leading-order contribution reads
\begin{equation}
\langle \gamma\rangle=\frac1{8v^2(\omega^2-{\cal V})}\int dx \langle v(x)v(0)\rangle \cos\left(\frac{2x\sqrt{\omega^2-{\cal V}}}v\right).
\label{eq:lyap-env-sol}
\end{equation}
Using Eq.~\eqref{eq:correlator} for the correlator $\Braket{v(x)v(0)}$, we note that that Eq.~\eqref{eq:lyap-env-sol} reproduces the semiclassical result of Sec.~\ref{sec:semicalss}, $\langle\gamma\rangle=(\Omega/\omega)^2/2\xi$ at $\omega\gg\sqrt{\cal V}\sim \Omega$.

\begin{figure}
\flushleft{(a)}\\
\centering
\includegraphics[width=0.9\columnwidth]{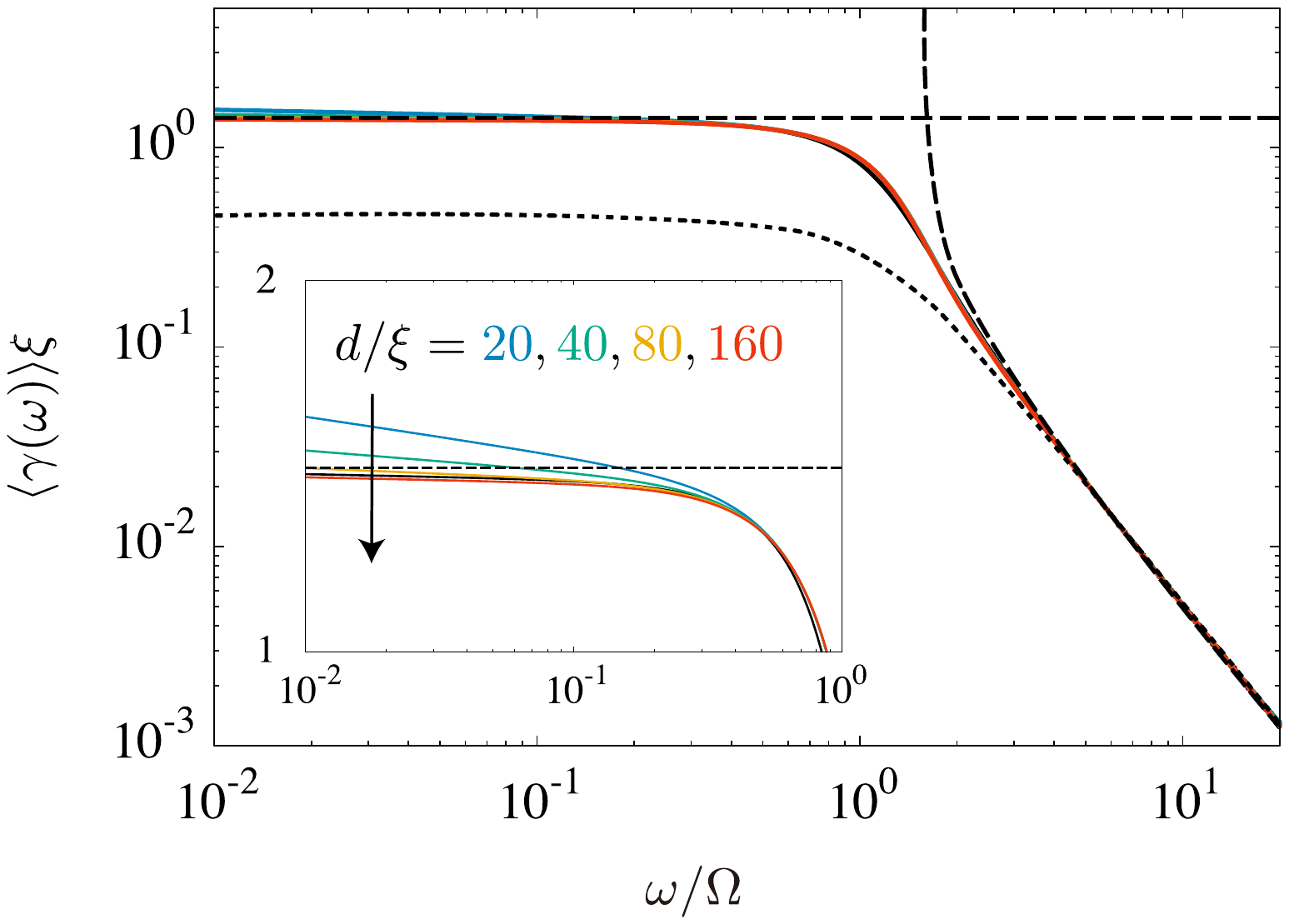}
\\
\flushleft{(b)}\\
\centering
\includegraphics[width=0.9\columnwidth]{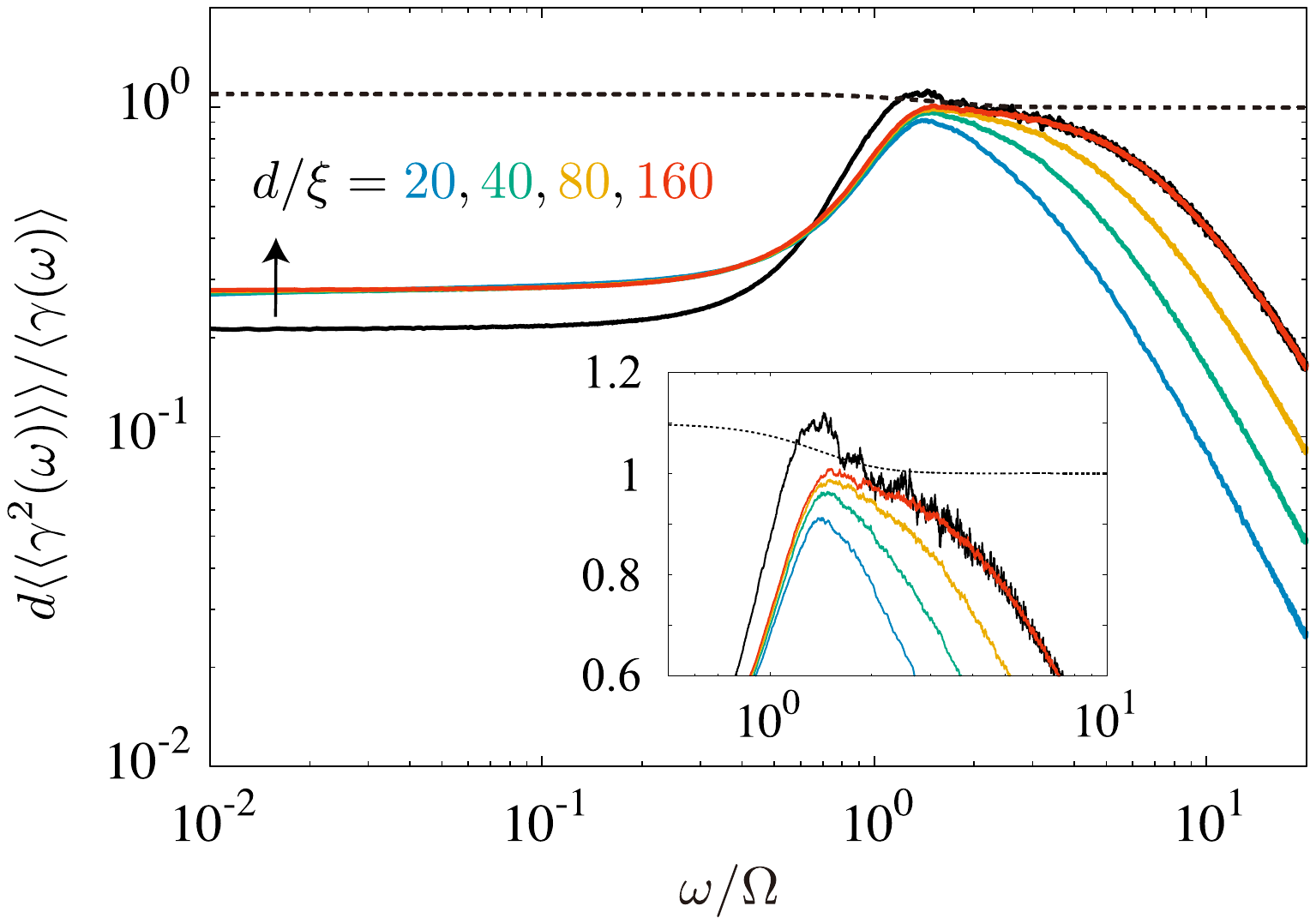}
\caption{\label{fig:lyapunov-gauss}
(a) The average of the Lyapunov exponent and (b) its variance as a function of frequency for different lengths $d/\xi=20,40,80,160$ and $K/K_0=1.0$ in the Gaussian colored model. The inset is an enlarged view of the low (a) and intermediate (b) frequency regions.
{The dotted lines in (a) and (b) show the result of the Gaussian white{-}noise potential at $d\to\infty$ for comparison, the black line shows the result for the pinned potential at $d/\xi=160$. The dashed lines in (a) show the result of the weak-disorder expansion for $\langle \gamma(\omega)\rangle$ at small and large $\omega$, Eqs.~\eqref{eq:Lyap-pert} and \eqref{eq:lyap-env-sol}, respectively. The Lyapunov exponents in the pinned and Gaussian colored models hardly differ from each other; we ascribe their similarity to the success of the weak-disorder expansion in most of the frequency range, apart from the close vicinity of $\Omega$. By contrast, the variances of the Lyapunov exponent computed in the Gaussian and pinned models differ from each other in the strongly localized regime}}
\end{figure}

The two asymptotes, Eqs.~\eqref{eq:Lyap-pert} and \eqref{eq:lyap-env-sol}, are shown in Fig.~\ref{fig:lyapunov-gauss}(a).

\subsection{Numerics}
\label{Gauss-num}

\begin{figure}
\includegraphics[width=0.9\columnwidth]{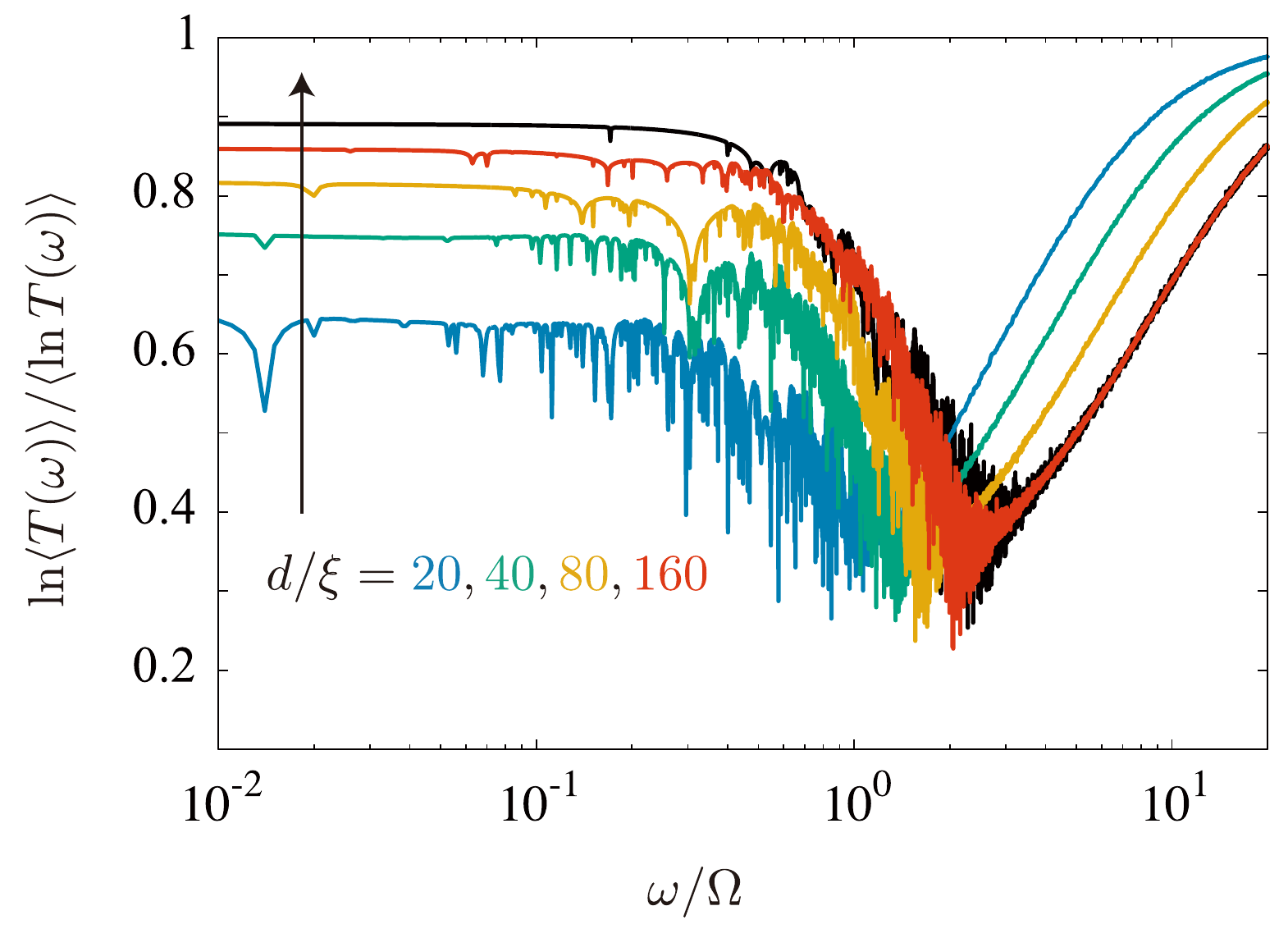}
\caption{\label{fig:ratio-lnT-gauss}
The frequency dependence of the ratio of $\ln\Braket{T(\omega)}$ and $\Braket{\ln T(\omega)}$ for different lengths $d/\xi=20,40,80,160$ and $K/K_0=1.0$ in the Gaussian colored model. A scaling analysis with the length {(see footnote~\cite{scaling})} gives the result $\ln\Braket{T(\omega)}/\Braket{\ln T(\omega)}\approx 0.87$ at {$\omega\ll\Omega$ and} $d/\xi \to \infty$. The analysis for the pinned model, illustrated with the black line corresponding to $d/\xi=160$, gives $\ln\Braket{T(\omega)}/\Braket{\ln T(\omega)}\approx 0.90$ at {$\omega\ll\Omega$ and} $d/\xi \to \infty$.} 
\end{figure}

\begin{figure}
\includegraphics[width=0.9\columnwidth]{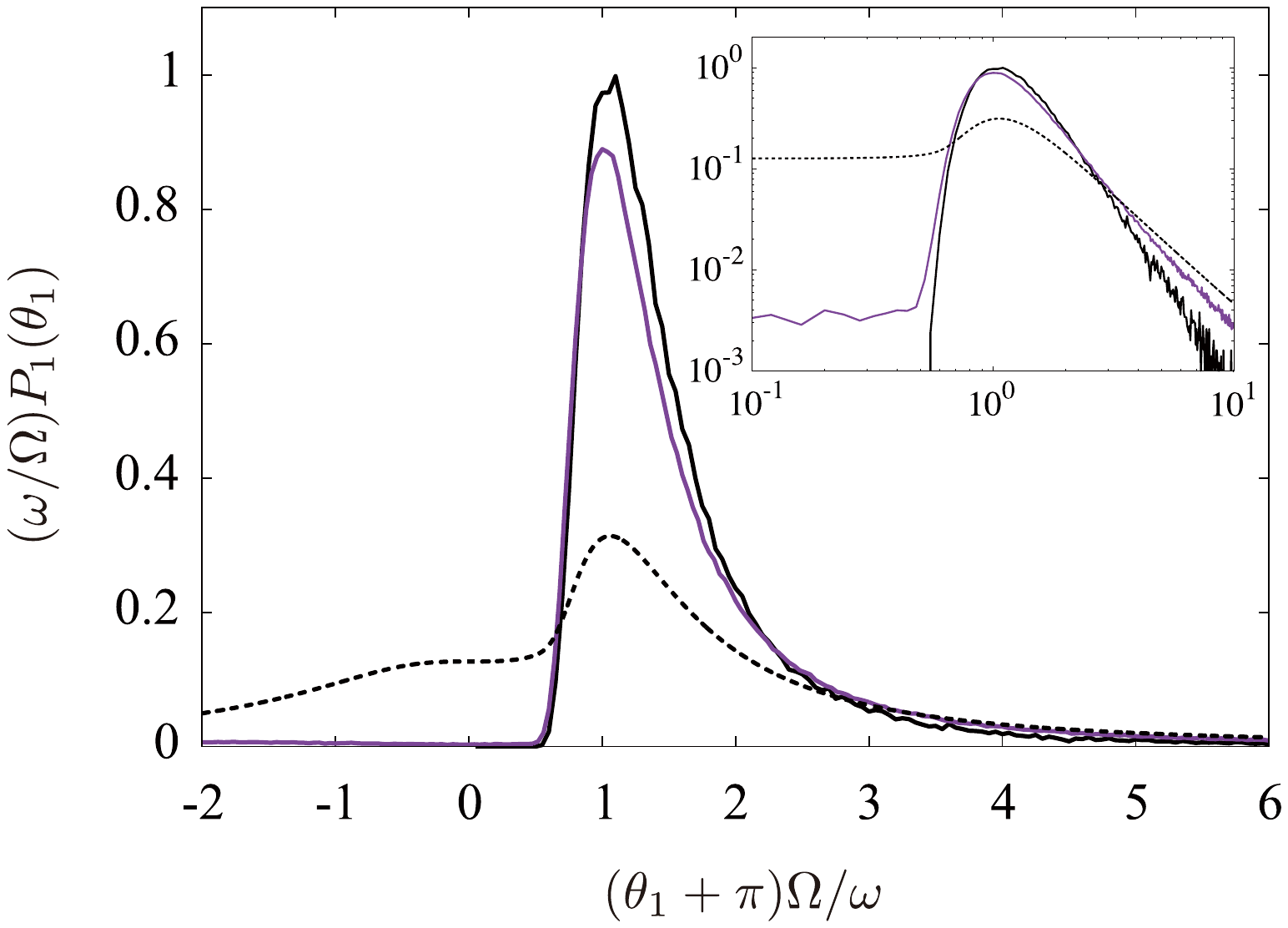}
\caption{\label{fig:proba-reflection-gauss}
Probability distributions of the reflection phase at $K/K_0=1.0$ and $d/\xi=20$, at $\omega/\Omega=0.1$, for the  Gaussian colored (purple), and pinned (black) models. The dotted line represents the analytical formula derived in Ref.~\cite{Sulem1973} for a Gaussian white{-}noise disorder. The inset illustrates the saturation of $P_1(\theta_1)$ to a finite value at $\theta_1=-\pi$ in the Gaussian models; by contrast, $P_1(\theta_1=-\pi)$ vanishes in the pinned model.} 
\end{figure}

\begin{figure}
\includegraphics[width=0.9\columnwidth]{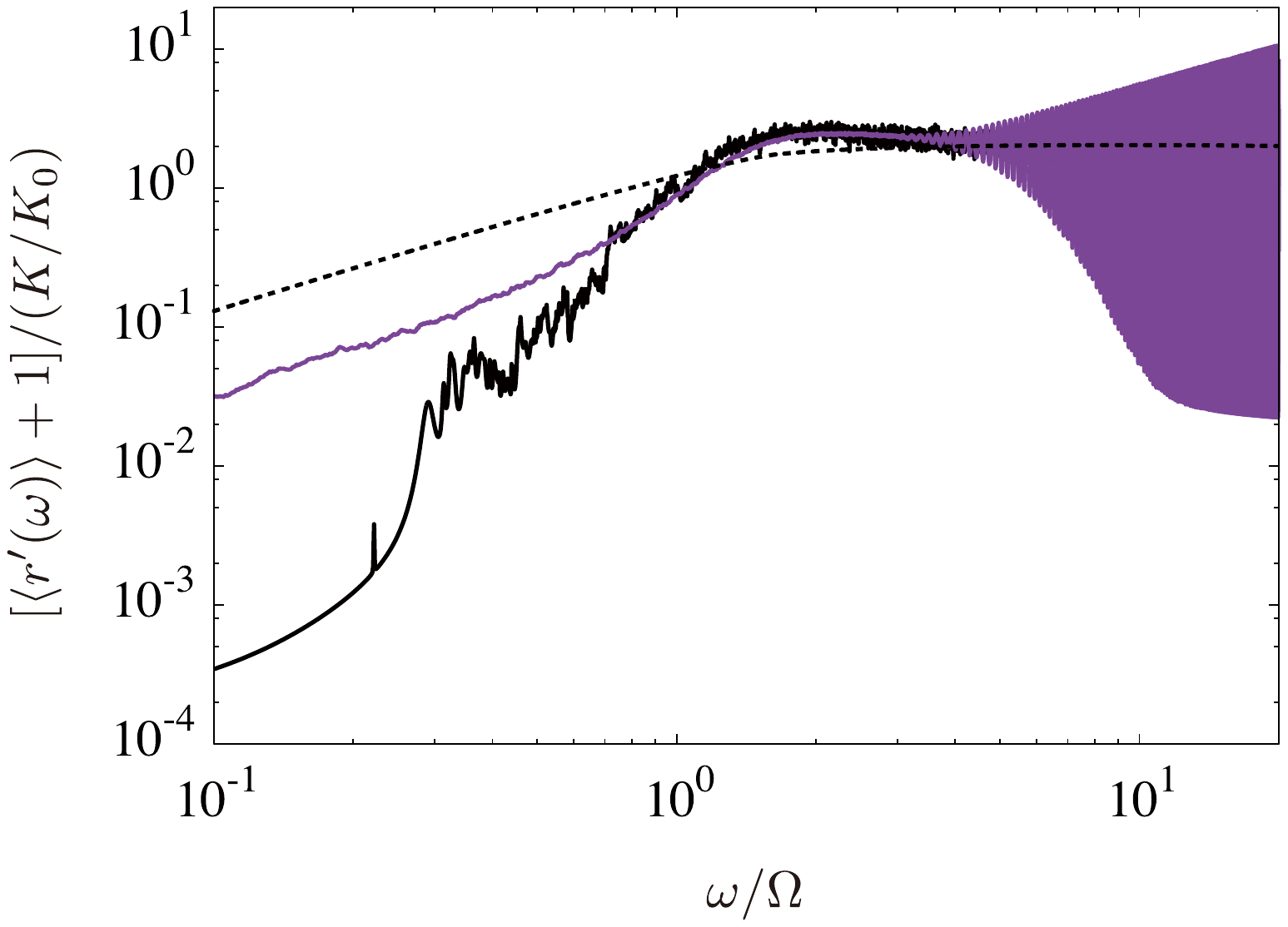}
\caption{\label{fig:DOS-gauss}
The local density of states as a function of the frequency for $K/K_0=0.01$.  The black and purple lines show the result of the pinned and Gaussian colored models, respectively. They differ in their frequency dependence, which is $\propto \omega^4$ in the first case and $\propto \omega$ in the second case, at $\omega\ll\Omega$. The dotted line is the result of the Gaussian white{-}noise model; it also shows the dependence $\propto \omega$ at $\omega\ll\Omega$.}
\end{figure}

We readily check that
\begin{equation}
{\cal V}_1(x)=\Braket{\cal V}+{\cal V}_0(x)+\int dy{\cal W}(x-y){\cal V}_0(y),
\end{equation}
where ${\cal V}_0$ is a Gaussian white-noise potential such that $\Braket{{\cal V}_0(x)}=0$ and $\Braket{{\cal V}_0(x){\cal V}_0(x)}=4\Omega^4\xi\delta(x-y)$, and 
\begin{subequations}
\begin{eqnarray}
{\cal W}(x)&=&\int \frac{dq}{2\pi}e^{iqx}\left[\sqrt{1-\frac{w_q}4}-1\right],\\
\quad w_q&=&\xi^{-1}\int dxe^{-iqx}w(x/\xi),
\end{eqnarray}
\end{subequations}
is a Gaussian colored potential that has the same average and second cumulant, Eq.~\eqref{eq:correlator}, as the pinned potential ${\cal V}(x)$. By implementing numerically the white noise potential ${\cal V}_0(x)$ on a lattice, as in Appendix~\ref{app:numerics}, we may then study the scattering properties of a medium characterized by the Gaussian colored potential ${\cal V}_1(x)$. In Fig.~\ref{fig:lyapunov-gauss}, we compare the frequency dependence of the Lyapunov exponent and its variance in the pinned and Gaussian colored models. Figures~\ref{fig:ratio-lnT-gauss}, \ref{fig:proba-reflection-gauss}, and \ref{fig:DOS-gauss} show the comparison for the ratio $\ln\Braket{T(\omega)}/\Braket{\ln T(\omega)}$, the probability distribution of the reflection phase, and the local density of states, respectively. They mirror Figs.~\ref{fig:lyapunov}, \ref{fig:ratio-lnT}, \ref{fig:proba-reflection}, and \ref{fig:DOS}, which were shown in the main text, see Sec.~\ref{sec:localized} for the discussion of the results.

\begin{thebibliography}{999}
\bibitem{footnote-reflection}
We evaluate $A=\int d\varphi p(\varphi)\varphi^2$ where $p(\varphi)$ is the numerically obtained curve plotted in the inset of Fig.~\ref{fig:proba-reflection}.
\bibitem{scaling}
For each length $d$, we find a linear fit of the low frequency {asymptotes, $\Braket{\ln T(\omega)}=(\omega/\Omega)A+B$ and ${\ln \Braket{T(\omega)}}=(\omega/\Omega)A'+B'$}. The numerics is well described by assuming that the slopes $A,A'$  scale inversely with $d/\xi$, and the intercepts $B,B'$ are the sum of a $d$-independent term and a term that scales inversely with $d/\xi$. The zero-frequency result at $d\to\infty$, which is quoted in the text, is the ratio of the $d$-independent terms of the intercepts.
\end{thebibliography}
\end{document}